\documentclass[letterpaper, 10 pt, conference]{ieeeconf}

\IEEEoverridecommandlockouts   

\usepackage{amsmath, amssymb, amsfonts, mathtools}  
\DeclareMathOperator{\sgn}{sgn}
\usepackage{theorem}
\usepackage{epsfig}
\usepackage[mathscr]{eucal}
\usepackage{cases}
\usepackage{bm,bbm,dsfont}
\usepackage{xcolor,ifpdf}

\usepackage{graphics,epstopdf} 
\usepackage{graphicx}           

\let\labelindent\relax
\usepackage{enumerate}
\usepackage[shortlabels]{enumitem}
\setlist[description]{leftmargin=\parindent}
\setlist[itemize]{leftmargin=*}

\usepackage{float}
\usepackage{subfigure}
\usepackage{soul}

\usepackage{tikz}
\usetikzlibrary{arrows,graphs} 
\usetikzlibrary{calc}

\newtheorem{theorem}{Theorem}
\newtheorem{definition}{Definition}
\newtheorem{lemma}{Lemma}
\newtheorem{corollary}{Corollary}
\newtheorem{proposition}{Proposition}

{\theorembodyfont{\upshape}
	\newtheorem{remark}{Remark}
	\newtheorem{example}{Example}
	
}

\newcommand{\bite}{\begin{itemize}}
	\newcommand{\eite}{\end{itemize}}
\newcommand{\benu}{\begin{enumerate}}
	\newcommand{\eenu}{\end{enumerate}}

\newcommand{\beq}{\begin{equation}}
	\newcommand{\eeq}{\end{equation}}
\newcommand{\beqa}{\begin{eqnarray}}
	\newcommand{\eeqa}{\end{eqnarray}}
\newcommand{\beqan}{\begin{eqnarray*}}
	\newcommand{\eeqan}{\end{eqnarray*}}

\newcommand{\real}[1]{\mathrm{Re}[#1]}
\newcommand{\epos}{\overset{\vee}{>}}

\newcommand{\Real}{\mathbb{R}}

\def\diag{{\rm diag}}

\def\bfone{{\bf 1}}

\newcommand{\prooftx}[1]{\noindent\textbf{Proof of #1.} }

\def\qed{\hfill \vrule height 5pt width 5pt depth 0pt \medskip}

\title{\LARGE \bf  Multi-agent consensus over time-invariant and time-varying signed digraphs via eventual positivity}

\author{Angela Fontan$^{1}$, Lingfei Wang$^{2}$, Yiguang Hong$^{3}$, Guodong Shi$^{4}$, and Claudio Altafini$^{5}$
	\thanks{*Work supported in part by grants from the Swedish Research Council (grant n. 2020-03701 to C.A.), the ELLIIT framework at Link\"oping, the Shanghai Municipal Science and Technology Major Project (Grant 2021SHZDZX0100 to Y.H.)
		and the National Natural Science Foundation of China (Grant 61733018 to Y.H.).}
	\thanks{$^{1}$Angela Fontan was with the Division of Automatic Control, Department of Electrical Engineering, Link\"{o}ping University, SE-58183 Link\"{o}ping, Sweden, and is now with the Division of Decision and Control Systems, KTH Royal Institute of Technology, SE-100 44 Stockholm, Sweden {\tt\small angfon@kth.se}}%
	\thanks{$^{2}$Lingfei Wang is with the Key Laboratory of Systems and Control, Academy
		of Mathematics and Systems Science, Chinese Academy of Sciences,
		and also with the University of Chinese Academy of Sciences, Beijing
		100190, China 
		{\tt\small wlf@amss.ac.cn}}%
	\thanks{$^{3}$Yiguang Hong is with the Department of Control Science and Engineering, Tongji University, Shanghai 201804, China {\tt\small yghong@iss.ac.cn}}%
	\thanks{$^{4}$Guodong Shi is with the Australian Center for Field Robotics, School of Aerospace, Mechanical and Mechatronic Engineering, The University of Sydney, NSW 2008, Sydney
		{\tt\small guodong.shi@sydney.edu.au}}
	\thanks{$^{5}$Claudio Altafini is with the Division of Automatic Control, Department of Electrical Engineering, Link\"{o}ping University, SE-58183 Link\"{o}ping, Sweden 
		{\tt\small claudio.altafini@liu.se}}
}

\begin{document}
	
	\maketitle

	\begin{abstract}
		Laplacian dynamics on signed digraphs have a richer behavior than those on nonnegative digraphs.
		In particular, for the so-called ``repelling'' signed Laplacians, the marginal stability property (needed to achieve consensus) is not guaranteed a priori and, even when it holds, it does not automatically lead to consensus, as these signed Laplacians may loose rank even in strongly connected digraphs.
		Furthermore, in the time-varying case, instability can occur even when switching in a family of systems each of which corresponds to a marginally stable signed Laplacian with the correct corank.
		In this paper we present conditions guaranteeing consensus of these signed Laplacians based on the property of eventual positivity, a Perron-Frobenius type of property for signed matrices. 
		The conditions cover both time-invariant and time-varying cases. 
		A particularly simple sufficient condition valid in both cases is that the Laplacians are normal matrices. 
		Such condition can be relaxed in several ways. For instance in the time-invariant case it is enough that the Laplacian has this Perron-Frobenius property on the right but not on the left side (i.e., on the transpose). 
		For the time-varying case, convergence to consensus can be guaranteed by the existence of a common Lyapunov function for all the signed Laplacians. 
		All conditions can be easily extended to bipartite consensus.
	\end{abstract}

	\section{INTRODUCTION}
	
	Distributed algorithms for computation and control on networks often rely on a Laplacian-like dynamics to achieve their goal. 
	The underlying assumption that is normally made is that the adjacency matrix of the graph is nonnegative, meaning that the agents collaborate to achieve a common goal.  
	In several applications, however, assuming that the adjacency matrix has nonnegative weights is a limitation.
	These include social networks, where the individuals can be ``friends" or ``enemies'', computer networks, where computers can trust or mistrust each other, and robot networks, where robots can collaborate or compete to accomplish a task. 
	More generally, in all contexts in which antagonism plays a role, it is more appropriate to assume that the weights of the graph can have both positive and negative values, i.e., to consider signed graphs \cite{Altafini2013Consensus,shi2019dynamics}. 
	Other contexts in which signed graphs appear include e.g. small-disturbance angle stability of microgrids \cite{Song2018Network,9137633}.

	When a graph is signed, there is more than one way to construct a {\em signed Laplacian} matrix.
	In particular, the two main alternatives that have been studied in the literature differ in how the diagonal elements are computed from the signed adjacency matrix.
	In the terminology of \cite{shi2019dynamics}, they are denoted ``opposing'' and ``repelling'' Laplacian. 
	In this paper we focus on the ``repelling'' Laplacian, whose main feature is  that it always has $0$ as an eigenvalue, but it may fail to be stable \cite{bronski2014spectral}. 
	Another complication that arises for ``repelling'' Laplacians is that strong connectivity of the graph no longer guarantees that the Laplacian has corank 1, meaning that even marginally stable ``repelling'' Laplacians may fail to lead to consensus when their kernel has dimension larger than 1.
	In the time-invariant case, conditions for stability are provided in \cite{Zelazo2014Definitness,Chen2016Definitness,Zelazo17Robustness,9137633} for signed undirected graphs, while for signed digraphs some partial results appear in \cite{Altafini2015Predictable,altafini2019investigating,shi2013agreement}. 
	In the time-varying case, we are not aware of any systematic study (unlike for the ``opposing'' signed Laplacian, for which an abundant literature exists \cite{Proskurnikov2014Consensus,Xia16Structural,Liu2017Exponential,meng2018uniform}). The only somewhat related paper we know is \cite{Jiang2019Output}, which however investigates a different problem, the so-called signed consensus. 
	The interesting aspect that appears when trying to solve the associated time-varying signed ``repelling'' consensus problem  (approximated as a system switching among a certain number of Laplacian matrices) is that stability can be lost even if the time-varying signed Laplacian is marginally stable and with corank 1 for all times. 
	This is in stark contrast to what happens in the ``nonnegative'' and ``opposing''  Laplacian cases, and more akin to what happens on ``ordinary'' (i.e., non Laplacian) time-varying linear systems. 
	In the nonnegative digraph case, in fact, a time-varying Laplacian never diverges, even though it may not converge \cite{Blondel2005Convergence}.
	Time-varying ``opposing'' signed  Laplacians, on the other hand, converge to zero as soon as one of the switching systems is not structurally balanced \cite{Proskurnikov2014Consensus,Xia16Structural,Liu2017Exponential}.
	In ordinary linear systems, instead, stability of all matrices of a switching system does not imply stability of the switching system \cite{liberzon2003switching,lin2009stability}, and divergence can occur for certain switching patterns. 
	The analysis that must be performed on our systems is therefore qualitatively different from that required for ``nonnegative'' and ``opposing'' time-varying Laplacians (which can never become unstable), as well as from that required for ``ordinary'' (i.e., not Laplacian) switching linear system (where each matrix is asymptotically stable, while Laplacians are only marginally stable).

	The twofold aim of this paper is 
	\begin{enumerate}
	    \item to provide a thorough stability analysis of signed Laplacians on digraphs in both time-invariant and time-varying cases, and
	    \item to completely solve the consensus problem for signed digraphs in both cases.
	\end{enumerate}  
The conditions we develop rely on Perron-Frobenius (PF) theory.
	If the canonical way of formulating the PF theorem (for the positive orthant) is to consider a matrix which is nonnegative or Metzler, it has been shown in \cite{Johnson2004Matrices,Noutsos2006Perron,Noutsos2008Rechability} that the category of matrices enjoying a PF property (namely, spectral radius which is a simple strictly dominating real eigenvalue of the matrix, of positive eigenvector) is strictly larger than nonnegative (or Metzler) matrices, and includes also matrices having some off-diagonal entries that are negative. 
	These matrices are called {\em Eventually Positive} (EP) because their powers become positive matrices after a certain exponent.
	They are called {\em Eventually Exponentially Positive} (EEP) when the matrix exponential becomes positive after a certain time.
	Eventual positivity has been used by  some of us to study consensus-like problems in \cite{Altafini2015Predictable,altafini2019investigating}, to represent linear systems which are externally but not internally positive in \cite{Altafini2016Minimal}, and to study Laplacian pseudoinverses in \cite{fontan2021properties}. 
	
	In this paper we show that indeed EEP matrices can provide an effective way to obtain consensus conditions for signed Laplacians. 
	If in the undirected case necessary and sufficient conditions can be obtained \cite{altafini2019investigating,Chen2016Definitness,Chen2016Characterization,Zelazo2014Definitness}, in the directed case (which has so far only been investigated with these tools in \cite{altafini2019investigating}), only sufficient conditions can be found. 
	The gap between necessity and sufficiency admits a neat interpretation: it corresponds to matrices that obey a ``right PF property'' but not a ``left PF property'', i.e., such that their transpose fails to satisfy the PF property.  
	We also show that for weight balanced digraphs our EEP conditions become necessary and sufficient, and that for normal Laplacians they become equivalent to positive semidefinitness of the symmetric part of the Laplacian.

	The same notions can be used also in the time-varying case. 
	What is shown in the paper is that a set of matrices that are simultaneously EEP and normal forms a consensus set, i.e., a set such that any switching sequence of matrices from the set (with arbitrary switching times) leads to consensus. 
	A straightforward consequence is that any set of EEP signed Laplacians on undirected graphs is always a consensus set, and convergence to consensus is always guaranteed.
	That is not true for digraphs in which the signed Laplacians are not normal matrices. 
	In this case, divergence can occur for certain switching patterns, as can be easily shown in examples.
	
	In the paper it is shown that the normality condition is sufficient but not necessary. 
	As a matter of fact, in the time-varying case, normality corresponds to all Laplacians admitting a Common Lyapunov Function (CLF) of quadratic type, in which the matrix associated to the quadratic form is equal to the identity. 
	The class of time-varying Laplacians achieving consensus can be extended considerably if we relax the normality assumption and allow for more general CLF. 
	In particular we show how to check the existence of general quadratic CLFs and also of CLFs which are homogeneous polynomials \cite{Chesi1} using Linear Matrix Inequalities (LMI). 
	In order to do so, we need to adapt the methods normally used for families of Hurwitz systems to families of marginally stable systems with the right corank. Unlike e.g. \cite{Meng2017Stability,Valcher2016Stability}, however, the focus is here on consensus, rather than uniform asymptotic stability. For that it is necessary to project the LMIs onto the orthogonal complement of the ``agreement subspace'', where uniform asymptotic stability tests can then be applied.

	The approach can be applied also to discrete-time systems, and to discrete-time consensus problems on signed digraphs.
	In fact, when a matrix has negative entries but still has row sums equal to 1, its marginal Schur stability is no longer guaranteed \cite{Eger2016Limits}. 
	In this case, the equivalent concept of {\em Eventually Stochastic} (ES) matrix can be defined and used in an analogous way. 
	In particular, the LMI-based results we obtain in this case complement those obtained for discrete-time switching systems containing mixtures of asymptotically stable and marginally stable modes \cite{Formasini2011Stability,Meng2017Stability,ZHENG2018294}, which normally deal with positive systems only and with uniform asymptotic stability (here we are instead interested in consensus).
	
	Finally, the approach can be generalized also to bipartite consensus, i.e., a form of consensus in which all agents converge to the same value in modulus but not in sign \cite{Altafini2013Consensus}.
	Unlike for the ``opposing'' signed Laplacian, where this concept is tightly linked to the notion of structural balance of the signed digraph \cite{Altafini2013Consensus}, here bipartite consensus has to do instead with the signature of the right PF eigenvector, and can occur also when the signed graph is structurally unbalanced.
	
	The rest of this paper is organized as follows: in Section \ref{sec:pre} some preliminary material is presented. The time-invariant case is studied in Section~\ref{sec:TI}, and the time-varying case in Section~\ref{sec:TV}. Section~\ref{sec:CLF} deals with CLF in the time-varying case and Section~\ref{sec:bipartite} with bipartite consensus. 
	
	Part of the material in this paper was presented in the two conference papers \cite{altafini2019investigating} and \cite{Wang22Multiagent}. In particular, a large fraction of the material of Section~\ref{sec:TI} appears in  \cite{altafini2019investigating}, and that of Section~\ref{sec:TV} in \cite{Wang22Multiagent}. 
	The material in Sections~\ref{sec:CLF} and~\ref{sec:bipartite} is novel and appears here for the first time.

	
	\section{Preliminary material}
	\label{sec:pre}
	
	\noindent $Notations.$ The real number and integer sets are denoted ${\mathbb{R}}$ and $\mathbb{Z}$, respectively, while $\mathbb{R}_\geq$ and $\mathbb{Z}_\geq$ represent respectively the nonnegative real number and nonnegative integer sets. In general, numbers are denoted lowercase letters $x,y,a,b,\dots$ and lowercase Greek letters $\alpha, \beta,\dots$ The modulus of a number is denoted $|\cdot|$. For any $x\in{\mathbb{R}}$, we write $\sgn(x)$ to denote the sign of $x$ and let $\sgn(0)=0$. Given a positive integer $m$, let $[m]$ be the set of all positive integers that are no larger than $m$, i.e., $[m]=\{1,2,\dots,m\}$. All vectors are real column vectors denoted with bold lowercase letters $\bf{x},\bf{y},\bf{z},\dots$ The $i$-th entry of a vector $\mathbf{x}$ is denoted $[\mathbf{x}]_i$. The Euclidean norm is denoted $\|\cdot\|$. Matrices are denoted with upper case letters such as $A,B,W, \dots$ All matrices are real unless stated otherwise. Given a matrix $A\in\mathbb{R}^{n\times n}$, $A^\top $ denotes its transpose and $A^k$ denotes the $k$th power of $A$. The $(i,j)$-th entry of a matrix $A$ is denoted $A_{ij}$ or $[A]_{ij}$; the spectrum of $A$ is denoted $\Lambda(A)=\{\lambda_1(A),\dots,\lambda_n(A)\}$, 
	while $\real{\lambda_i(A)}$ indicates the real part of the eigenvalue $\lambda_i(A)$; $\rho(A)$ represents the spectral radius of $A$, i.e., $\rho(A)=\max_{i\in [n]} |\lambda_i(A)|$. 
	The identity matrix is denoted $I_d$, with $d>0$ as the dimension (sometimes omitted). The vectors or matrices with all entries equal to $0$ or $1$ are all denoted $\mathbf{0}$ or $\mathbf{1}$, with the dimensions depending on the context.  
	Given $ A, \, B \in \mathbb{R}^{n \times n }$, their Kronecker product is denoted $ A \otimes B $ and their Kronecker sum $ A \oplus B $. 

	\subsection{Signed graphs and signed Laplacians}
	
	A digraph is represented as $\mathcal{G}=(\mathcal{V},\mathcal{E},A)$, where $ \mathcal{V} =[n]$ is an index set, an ordered pair $(j,i)\in \mathcal{E}$ denotes a directed link from node $j$ to node $i$ over the set $\mathcal{V}$, and the matrix $A\in\mathbb{R}^{n\times n}$ is the weighted adjacency matrix corresponding to $\mathcal{G}$, with $A_{ij}\not= 0$ if and only if $(j,i)\in\mathcal{E}$. Note that $A_{ij}$ can be positive or negative, which attaches to each edge $(j,i)$ a sign, i.e., $\sgn(A_{ij})$. The graph $\mathcal{G}$ is therefore called a \emph{signed graph}. $\mathcal{G}$ is undirected if $A=A^\top $. $\mathcal{G}$ is called weight balanced if $A\mathbf{1}=A^\top \mathbf{1}$.

	Given the signed digraph $ \mathcal{G}$ of weighted adjacency matrix $A$, and denoted $ \sigma^{\rm in}_i = \sum_{j=1}^n A_{ij} $ the weighted in-degree of node $i$, let us define the Laplacian as
	\beq
	L = \Sigma - A \qquad \text{where} \qquad \Sigma = \diag( \sigma^{\rm in}_1, \ldots, \sigma^{\rm in}_n ) .
	\label{eq:Laplacian1}
	\eeq
	Since $A$ can have negative entries, $ \sigma^{\rm in}_i $ can even become negative.
	In our recent paper \cite{shi2019dynamics}, the Laplacian \eqref{eq:Laplacian1} is referred to as ``repelling signed Laplacian'', terminology which allows us to distinguish it from a second signed Laplacian (referred to in \cite{shi2019dynamics} as ``opposing signed Laplacian''), obtained replacing $ \sigma^{\rm in}_i $ with $ \sigma^{\rm in, abs}_i = \sum_{j=1}^n | A_{ij} | $, see \cite{Altafini2013Consensus,Belardo2014Balancedness,Hou2003Laplacian}.

	For the graph $\mathcal{G}$, node $i$ is said to be linked to $j$ if there exists an edge sequence $(j,i_1),(i_1,i_2),\dots,(i_{s-1},i_s),(i_s,i)$ that is picked from $\mathcal{E}$. We call $\mathcal{G}$ strongly connected if each pair of nodes in $\mathcal{V}$ is linked to each other.
	A graph $ \mathcal{G} $ has a rooted spanning tree if all nodes are linked to $j$ for some $ j \in \mathcal{V}$.

	\subsection{Matrix theory}
	
	Given a square matrix $A\in\mathbb{R}^{n\times n}$,  we say that $A$ is Hurwitz stable (resp. Schur stable) if $\mathrm{Re}[\lambda_i(A)]<0$ (resp. $ | \lambda_i(A) | < 1 $) for any $i$, and it is marginally stable (resp. marginally Schur stable) if $\mathrm{Re}[\lambda_i(A)]\leq0$ (resp. $ | \lambda_i(A) | \leq 1 $) and $ \lambda_i (A) $ such that $\mathrm{Re}[\lambda_i(A)] = 0$ (resp. $ | \lambda_i(A) | = 1 $) is a simple root of the minimal polynomial of $A$. 
	A matrix $A\in\Real^{n\times n}$ is said to be irreducible if there does not exist a permutation matrix $P$ such that $P^\top AP$ is block triangular, that is
	$$
	P^\top AP\neq\left[\begin{array}{cc}
		A_{1} & A_{2}\\
		0 & A_{3}
	\end{array}\right]\,,
	$$
	where $A_{1}$ and $A_{3}$ are nontrivial square matrices.
	The digraph $ \mathcal{G} $ of adjacency matrix $A$ is strongly connected if and only if $ A $ is irreducible.

	$A\in\mathbb{R}^{n\times n}$ is said to be a positive matrix, denoted $ A>0$, (resp. nonnegative, denoted $ A\geq 0$) if $A_{ij}>0$ (resp. $A_{ij}\geq 0$) for all $i,j\in [n]$. 
	Given $ A \geq 0$, the matrix $B=sI-A$, $ s>0$, is called a Z-matrix.
	If in addition $ s \geq \rho(A) $, then $B$ is called an M-matrix. 
	In particular, an M-matrix $B$ in which $s>\rho(A)$ is nonsingular and such that $ -B$ is Hurwitz stable. 
	If instead $s=\rho(A)$, $B$ is a singular M-matrix.
	If in addition $ A$ is irreducible, then $ -B $ is also marginally stable.
	
	When $A\in \mathbb{R}^{n\times n }$, the comparison matrix of $A$, denoted $M(A)$, has $ |A_{ii} | $ on the diagonal and $ - | A_{ij} |$ in the entry $ (i,j)$, $ i\neq j$.
	A matrix $A$ is called an H-matrix if its comparison matrix $M(A)$ is an M-matrix. 
	It is said an $H_+$-matrix if in addition $ A_{ii}\geq 0$, $ i=1, \ldots, n$, \cite{Hershkowitz1985Lyapunov}. 
	
	$A\in \mathbb{R}^{n\times n }$ is said to have corank $d$ if the dimension of the kernel space of $A$, $ \ker (A)$, is $d$. $A$ is called normal if $AA^\top  = A^\top  A$.
	A matrix $ A $ is said range symmetric \cite{Meyer2000} if $\ker (A)= \ker(A^\top)$ (and hence ${\rm range}(A)={\rm range}(A^\top)$).

	If $A\in \mathbb{R}^{n\times n }$ is symmetric, it is called positive definite (pd) if  $\mathbf{x}^\top  A\mathbf{x}> 0$  for all $\mathbf{x}\in\mathbb{R}^n$, $ \mathbf{x}\neq 0$, and positive semidefinite (psd) if $\mathbf{x}^\top  A\mathbf{x}\geq 0$ for all $\mathbf{x}\in\mathbb{R}^n$. A  pd (resp. psd) matrix $A$ is sometimes indicated $ A \succ 0 $ (resp. $ A \succeq 0$).

	\subsection{Perron-Frobenius property and eventual positivity}
	
	\begin{definition}[Perron-Frobenious property]\label{def:pf}
		A matrix $A\in\mathbb{R}^{n\times n}$ is said to have the (strong) Perron-Frobenious (PF) property, denoted $A\in\mathcal{PF}_n$, if $\rho(A)$ is a simple real positive eigenvalue of $A$ such that $\rho(A)>|\lambda|$ for all $\lambda \in \Lambda(A)$, $\lambda\ne \rho(A)$, and the corresponding right eigenvector is positive.
	\end{definition}
	
	\begin{definition}[Eventually positive]
		A matrix $A\in\mathbb{R}^{n\times n}$ is called Eventually Positive (EP) if there exists $t_0\in\mathbb{Z}_{\geq}$ such that $A^t$ is positive for all $t\geq t_0$.
	\end{definition}
	Following \cite{Olesky2009Mv-matrices}, EP matrices will be denoted $ A \stackrel{\vee}{>} 0 $.
	The following necessary and sufficient condition relates EP matrices and PF property. 
	\begin{theorem} 
		\label{thm:PFforEvPos}
		(\cite{Noutsos2006Perron}, Theorem~2.2) 
		For $ A \in \mathbb{R}^{n\times n}$ the following are equivalent:
		\begin{enumerate}
			\item Both $ A , \, A^\top  \in \mathcal{PF}_n $;
			\item $ A\stackrel{\vee}{>} 0 $;
			\item $ A^\top \stackrel{\vee}{>} 0 $.
		\end{enumerate}
	\end{theorem}
	Also the following lemma will be useful later on.
	\begin{lemma}
		\label{lemma:no_two_pos-eig}
		(\cite{Altafini2015Predictable}, Lemma 1)
		Consider $A \stackrel{\vee}{>} 0$ and denote $ {\bf v}>0$ its right PF eigenvector. 
		Then any eigenvector $ {\bf v}_1 $ of $A$ such that $ {\bf v}_1 >0$ must be a multiple of $ {\bf v}$. 
	\end{lemma}
	
	Recall that a matrix $ A\in \mathbb{R}^{n\times n}$ is said exponentially positive if $ e^{At} = \sum_{k=0}^\infty \frac{A^k t^k}{k!} >0$ $ \forall \, t\geq 0$, and that $ A$ is exponentially positive if and only if $ A_{ij} \geq 0 $ $ \forall \, i \neq j$ \cite{Noutsos2008Rechability}. 
	\begin{definition}[Eventually exponentially positive]\label{def:ev_exp_pos}
		A matrix $A\in\mathbb{R}^{n\times n}$ is called Eventually Exponentially Positive (EEP) if there exists $t_0\ \in \mathbb{R}_\geq $ such that $e^{At}$ is positive for all $t\geq t_0$.
	\end{definition}
	The relationship between EP and EEP is provided by the following lemma.
	\begin{lemma}
		\label{lemma:ev_exp_pos}
		(\cite{Noutsos2008Rechability}, Thm~3.3)
		A matrix $ A \in \mathbb{R}^{n\times n}$ is EEP if and only if $ \exists $ $ d \in \mathbb{R}_\geq 0 $ such that $ A + d I \stackrel{\vee}{>} 0$.
	\end{lemma}

	\subsection{Eventually stochastic matrices}
	Recall that a matrix $W$ is row (resp. column) stochastic if $ W \mathbf{1} = \mathbf{1}$ (resp. $ \mathbf{1}^\top  W =  \mathbf{1}^\top  $),  $ 0 \leq w_{ij} \leq 1 $, and doubly stochastic if it is both row and column stochastic.

	\begin{definition}[Eventually stochastic]
		A matrix $W\in\mathbb{R}^{n\times n}$ is called Eventually Stochastic (ES) if $W$ is EP and $W\mathbf{1}=\mathbf{1}$. If, moreover, $W^\top  \mathbf{1}=\mathbf{1}$, $W$ is called Eventually Doubly Stochastic (EDS).
	\end{definition}
	
	The following lemma follows from the lemma in Sect. VI.C of \cite{Altafini2015Predictable}, and from Theorem~\ref{thm:PFforEvPos} and Definition~\ref{def:pf}.

	\begin{lemma}
		\label{lemma:ev_stoch}
		If $W$ is ES, then $ \rho(W) =1$ is a simple positive eigenvalue of $W$ such that  $\rho(W)>|\lambda|$ for all $\lambda \in \Lambda(W)$, $\lambda\ne \rho(W)$, and the corresponding right and left eigenvectors are positive.
	\end{lemma}
	
	%
	%
	%
	%
	%
	%
	
	\subsection{Signed Perron-Frobenius property}
	
	\begin{definition}[Signed Perron-Frobenious property]\label{def:spf}
		A matrix $A\in\mathbb{R}^{n\times n}$ is said to have the (strong) signed PF property, denoted $A\in\mathcal{SPF}_n$, if $\rho(A)$ is a simple real positive eigenvalue of $A$ such that $\rho(A)>|\lambda|$ for all $\lambda \in \Lambda(A)$, $\lambda\ne \rho(A)$, and the corresponding right eigenvector $ \mathbf{v}_r $ is such that $ | \mathbf{v}_r | >0$.
	\end{definition}
	
	The special case that interests us is when $ A , \, A^\top \in \mathcal{SPF}_n$ and both $ \mathbf{v}_\ell $ and $ \mathbf{v}_r $,  the left and right eigenvectors associated to $ \rho(A) $, have the same sign pattern.
	
	\begin{theorem}\label{thm:spf}(Proposition of Sect.~V of \cite{Altafini2015Predictable})
		Given $ A \in \mathbb{R}^{n\times n}$, then the following are equivalent:
		\begin{enumerate}[(i)]
			\item $A , \, A^\top \in \mathcal{SPF}_n$, and $ \mathbf{v}_\ell$, $ \mathbf{v}_r$ such that $ [\mathbf{v}_\ell]_i [\mathbf{v}_r]_i >0 $ $ \forall \, i=1, \ldots, n$, or $ [\mathbf{v}_\ell]_i [\mathbf{v}_r]_i <0 $ $ \forall \, i=1, \ldots, n$;
			\item $ \exists \, S = {\rm diag}(\mathbf{s} ) $, with $ \mathbf{s} = \begin{bmatrix} s_1 & \ldots & s_n \end{bmatrix}^\top $, $ s_i = \pm 1 $, such that $ SAS \epos 0$.
		\end{enumerate}
	\end{theorem}
	
	The matrix $ S ={\rm diag}(\mathbf{s}) $ in the previous theorem is called a diagonal signature matrix, of signature $ \mathbf{s}$. 
	For this special case we shall use the following special definitions.

	\begin{definition}
		A matrix $A\in\mathbb{R}^{n\times n}$ is said
		\begin{enumerate}
			\item Signed Eventually Positive (SEP) if $ \exists $ a diagonal signature matrix $ S $ such that $ SAS $ is EP;
			\item Signed Eventually Exponentially Positive (SEEP) if $ \exists $ a diagonal signature matrix $ S $ such that $ SAS $ is EEP;
			\item Signed Eventually Stochastic (SES) if $ \exists $ a diagonal signature matrix $ S={\rm diag}(\mathbf{s})  $ such that $ SAS $ is ES and $ A \mathbf{s} = \mathbf{s}$. 
		\end{enumerate}
	\end{definition}

	
	\section{Consensus on time-invariant signed digraphs}
	\label{sec:TI}
	
	\subsection{Problem formulation}
	Consider a signed digraph $ \mathcal{G}$ over $ \mathcal{V} $. The state vector of the agents at time $t$ is given by $\mathbf{x}(t)\in\mathbb{R}^n$, with $[\mathbf{x}(t)]_i$ assigned to agent $i$ for all $i\in\mathcal{V}$.  Consider the following two consensus protocols:
	
	\begin{enumerate}[1.]
		\item  \textbf{Continuous-time protocol}. If $L$ is the signed Laplacian \eqref{eq:Laplacian1} associated to $ \mathcal{G}$, the system we consider is 
		\begin{equation}
			\dot{\mathbf{x}} = - L \mathbf{x} \,.
			\label{eq:ode:laplacian}
		\end{equation}
		\item  \textbf{Discrete-time protocol}. Denote $ W $ the signed adjacency matrix associated to $ \mathcal{G}$, with the property that $ W \mathbf{1} = \mathbf{1}$. The discrete-time protocol is then
		\begin{equation}
			\mathbf{x}(t+1) = W \mathbf{x}(t).
			\label{eq:discrete_sys}
		\end{equation}
	\end{enumerate}
	
	\begin{definition}
		\label{def:cons}
		We say that the system \eqref{eq:ode:laplacian} or \eqref{eq:discrete_sys} achieves consensus if, for all $\mathbf{x}(0)\in\mathbb{R}^n$, there exists $\alpha\in\mathbb{R}$ such that $\mathbf{x}^\ast = \lim_{t\to\infty}\mathbf{x}(t)=\alpha \mathbf{1}$.
	\end{definition}
	
	\textbf{Problem of interest}:  find conditions on the signed $ L$ and $W$ that guarantee that the systems \eqref{eq:ode:laplacian} and \eqref{eq:discrete_sys} achieve consensus. 
	
	\medskip

	The presence of signs in $ L $ and $ W$ complicates things with respect to the unsigned case. For instance, for signed Laplacians $L$ we have the following easily verifiable properties (similar properties hold also in discrete-time).
	\begin{proposition}
		\label{prop:signed-prop}
		Consider a signed, strongly connected digraph $ \mathcal{G} $. Then for the corresponding Laplacian \eqref{eq:Laplacian1} we have 
		\begin{enumerate}
			\item $0$ is always an eigenvalue of right eigenvector $ \bfone$;
			\item The multiplicity of the $0$ eigenvalue can be $ \geq 1 $;
			\item $L$ need not be diagonally dominant;
			\item $-  L$ need not be marginally stable;
			\item The quadratic form $ x^\top  L x $ need not be nonnegative, i.e., $ (L+ L^\top)/2 $ need  not be psd.
		\end{enumerate}
	\end{proposition}
	The proof is in Appendix~\ref{app:TI}.
	
	As a consequence, $ L$ and $ W$ can become unstable, and no longer lead to consensus in \eqref{eq:ode:laplacian} and \eqref{eq:discrete_sys}.

	
	\subsection{Continuous-time case}
	\label{sec:continuous}
	
	From Proposition~\ref{prop:signed-prop}, when $ \mathcal{G} $ is a signed graph, $L$ may fail to be diagonally dominant (technically, $L$ need not be an H-matrix), which may or may not lead to loss of stability of $-L$, in a way which is subtle to check, especially since there is no longer a correspondence between irreducibility of the Laplacian and its corank (Property~2 of Proposition~\ref{prop:signed-prop}), as the following example shows.
	
	\begin{example}
		\label{ex:corank2}
		Consider a complete, undirected, signed graph $\mathcal{G}$ whose Laplacian is 
		\begin{equation*}
			L= \begin{bmatrix} 
				3&-1&-1&-1\\
				-1&1&1&-1\\
				-1&1&1&-1\\
				-1&-1&-1&3
			\end{bmatrix}.
		\end{equation*}
		It is $\Lambda(L)=\{0,0,4,4\}$ and $\mathbf{1}, \begin{bmatrix}0 & 1&-1& 0\end{bmatrix} ^\top $ are both eigenvectors in $ \ker(L)$, and $0$ has multiplicity $1$ in the minimal polynomial of $L$.  Hence $L$ is marginally stable of corank $2$. See also \cite{PanShaoMesbahi2016} for related observations.
		$\hfill\square$
	\end{example}
	
Also a loss of rank can lead to loss of consensus, even in presence of marginal stability. 
The next lemma highlights the key role of the corank of $L$, which for signed graphs replaces irreducibility.

	\begin{lemma}\label{le:nec}
		Given the time invariant system \eqref{eq:ode:laplacian} with $L$ a signed Laplacian matrix, consensus is achieved if and only if $-L$ is marginally stable of corank $1$.
	\end{lemma}
	The proof is in Appendix~\ref{app:TI}.

	\medskip
	
	\noindent {\bf Example~1 (cont'd)}
In this example it is $ \lim_{t\to\infty}   e^{-Lt} \neq \mathbf{1} \mathbf{c}^\top $, hence the system \eqref{eq:ode:laplacian} does not converge to consensus for this corank 2 Laplacian, in spite of marginal stability.
	$\hfill\square$
	
	\medskip
	
	Our task in the following is therefore to determine conditions that guarantee both the marginal stability and the correct corank of $ -L$.
	We first summarize the known results for undirected graphs, and then develop our new results for digraphs.


	\medskip
	
	\subsubsection{Signed undirected graph case}

	The case of $ \mathcal{G}$ signed and undirected has been studied extensively in the literature, mostly in terms of the so-called effective resistance matrix \cite{Chen2016Definitness,Chen2016Characterization,Zelazo2014Definitness}.
	Following \cite{altafini2019investigating}, here we express instead the condition of $L$ psd in terms of EEP matrices.

	\begin{theorem}
		\label{thm:signed-undirected-cont}
		Consider an undirected signed graph $ \mathcal{G}$ of Laplacian $ L $. $L$ is psd of $ {\rm corank}(L)=1$ (and hence marginally stable of $ {\rm corank}(L)=1$) iff $-L$ is EEP.
	\end{theorem}
	
	
	The proof can be found in \cite{altafini2019investigating}, Theorem~3, see also \cite{9137633} for related results.

	
	\medskip
	
	\subsubsection{Signed digraph case}
	\label{sec:signed_digraph}
	
	In the signed digraph case, the conditions we obtain are no longer necessary and sufficient for marginal stability of $-L$.

	%
	%
	
	\begin{theorem}
		\label{thm:signed-digraph1}
		Consider a signed digraph $\mathcal{G}$, and the corresponding Laplacian $L$. If $-L$ is EEP, then $-L$ is marginally stable of corank 1, and the system \eqref{eq:ode:laplacian} converges to 
		\[
		\mathbf{x}^\ast = \lim_{t\to\infty} \mathbf{x}(t) = \frac{{\bf v}_\ell ^\top  \mathbf{x}(0) \mathbf{1} }{{\bf v}_\ell ^\top  \mathbf{1} },
		\]
		where $ {\bf v}_\ell$ is the left eigenvector of $L$ relative to $ 0 $. 
		Viceversa, if $- L$ is 
		marginally stable and of corank 1 then $ \exists $ a scalar $ d\geq 0 $ such that $ dI-L \in \mathcal{PF}$.
	\end{theorem}
	The proof is in Appendix~\ref{app:TI}.

	\begin{remark}
		The gap between the two conditions of Theorem~\ref{thm:signed-digraph1} corresponds to matrices $L$ s.t. $ B = dI -L  \in \mathcal{PF}$ for some $ d\geq 0 $, but $ B^\top  \notin \mathcal{PF}$ for all $ d\geq 0$. 
		The stability of such class of Laplacians cannot be determined a priori, as the following two examples show. 
	\end{remark}
	
	\begin{example}
		\label{example-PF-notPF_stable}
		In correspondence of 
		\[
		L=\begin{bmatrix}          
			-0.4  &  0.7    &    0   & -0.3 \\
			-1.4  &    1.6  & 0.2  &  -0.4 \\
			-0.7  &     0   &      2.8   & -2.1 \\
			0    &     0   & -1.3   &   1.3\end{bmatrix}
		\]
		it is $ \Lambda (L) =\{ 0, \,  0.7325 \pm 0.1220i ,\, 3.8349 \}$, i.e. $-L$ is marginally stable, even though the left eigenvector associated to $0$ is not positive. For $ d>1.92 $, $B=dI - L \in \mathcal{PF}$ but $ B^\top  \notin \mathcal{PF}$.
		Notice also that $L$ has both positive and negative values on the diagonal. $\hfill\square$
	\end{example}
	
	\begin{example}
		\label{example-PF-notPF_unstable}
		For
		\[
		L=\begin{bmatrix} 
			1.4  &   -1.9 & 0.5   &      0 \\
			0   &      1     &    0  &  -1 \\
			0.3  & -0.2   &    0.1  &  -0.2 \\
			-1.8   &     0    &     0    &     1.8 \end{bmatrix}
		\]
		it is $ \Lambda (L) =\{ -0.0890, \, 0, \,  2.1945 \pm 1.2509i\}$, i.e. $- L$ is unstable. Clearly $ x^\top  L x $ is not psd. Also in this case the left eigenvector associated to $0$ is not positive, and, for $ d>2.6 $, $B=dI - L \in \mathcal{PF}$ but $ B^\top  \notin \mathcal{PF}$. $\hfill\square$
	\end{example}
	
	The following is instead an example of EEP matrix.
	
	\begin{example}
		\label{example4}
		For
		\[
		L=\begin{bmatrix}
			1  &   -1  &      0   &       0 \\
			0   &       2.6  &  - 2.6    &      0 \\
			-0.3    &     0   &       1.4  &   -1.1 \\
			-0.9  & 0.2  &   -0.9    &      1.6 \end{bmatrix}
		\]
		we have $ \Lambda (L) =\{0, \, 1.6956 \pm 0.9452i, \,   3.2089\}$. 
		For $ d\geq1.61 $ it is $ B = dI - L  \stackrel{\vee}{>} 0$.
		$\hfill\square$
	\end{example}
	
	In the case of $L$ weight balanced , however, there is no gap between necessity and sufficiency.
	
	\begin{corollary}
		\label{cor:weight-bal1}
		Consider a 
		signed digraph $\mathcal{G}$ such that the corresponding Laplacian \eqref{eq:Laplacian1} is weight balanced. 
		Then the following conditions are equivalent:
		\begin{enumerate}[(i)]
			\item $-L$ is EEP;
			\item $-L$ is marginally stable of corank 1;
		\end{enumerate}
		Furthermore, if $L$ is normal then (i) and (ii) are equivalent to
		\begin{enumerate} 
			\item[(iii)] $ L_s =(L+L^\top  )/2$ is psd of corank 1.
		\end{enumerate}
	\end{corollary}
The proof is in Appendix~\ref{app:TI}.

\begin{remark}
Corollary 2 of \cite{altafini2019investigating} erroneously claims that the equivalence (ii) $ \Longleftrightarrow $ (iii) in Corollary~\ref{cor:weight-bal1} is valid in the more general case of weight balance $L$, which is not true.
Only one direction is valid ((iii)$ \Longrightarrow$ (ii)), as shown in Corollary~\ref{lem:psd-EP}.
As for the other direction, a complication arises for instance from the fact that for $ L$ weight balanced but not normal $ L_s $ may acquire negative diagonal elements even if $ -L $ is marginally stable, see Example~\ref{example:Ls_nonpsd}. $ L_s $ with negative diagonal elements obviously cannot be psd. However, even when $ L_s $ has positive diagonal, it is not guaranteed to be psd, see Example~\ref{example:Ls_nonpsd_posdiag}. Normality of $L$ guarantees instead psd of its symmetric part, although it is not necessary, see Example~\ref{ex:norm-not-nec}.
\end{remark}
	

	
\begin{corollary}\label{lem:psd-EP} 
If $L_s$ is psd of corank $1$, then $L$ is weight balanced and marginally stable of corank $1$.
\end{corollary}
The proof is in Appendix~\ref{app:TI}.

	\begin{example}\label{example:Ls_nonpsd}
		In correspondence of 
		\begin{equation*}
			L=\begin{bmatrix} 
				0.15&0&0&-0.15\\
				-0.23&0.15&0.15&-0.07\\
				0.01&-0.12&-0.03&0.14\\
				0.07&-0.03&-0.12&0.08
			\end{bmatrix}
		\end{equation*}
		it is $\Lambda(L)=\{0, 0.0901\pm 0.199 i ,0.169\}$, i.e., $-L$ is marginally stable of corank $1$. Moreover, $L\mathbf{1} = L^\top\mathbf{1}=0$ and, for $d>0.2647$, $B= d I -L \epos 0$. 
		However, $\Lambda(L_s)=\{-0.0402, 0, 0.1248 ,0.2655\}$, i.e., $L_s$ is not psd.
	\end{example}
	
	\begin{example}\label{example:Ls_nonpsd_posdiag}
		For
		\begin{equation*}
			L=\begin{bmatrix} 
				0.23&0&-0.28&0.05\\
				-0.01&0.03&0.02&-0.04\\
				0.05&-0.03&0.04&-0.06\\
				-0.27&0&0.22&0.05
			\end{bmatrix}
		\end{equation*}
		it is $\Lambda(L)=\{0, 0.1443\pm 0.1859 i ,0.0514\}$, i.e., $-L$ is marginally stable of corank $1$. Moreover, $L\mathbf{1} = L^\top\mathbf{1}=0$ and, for $d>0.1919$, $B= d I -L \epos 0$. 
		However, $\Lambda(L_s)=\{-0.0446, 0, 0.0404 ,0.3441\}$, i.e., $L_s$ is not psd.
	\end{example}

	\begin{example} \label{ex:norm-not-nec}
	For
		\begin{equation*}
			L=\begin{bmatrix} 
				1&1&-1&-1\\ -1&1&0&0\\
				-1&-1&2&0\\  1&-1&-1&1
			\end{bmatrix},
		\end{equation*}
		which is not normal, it is $\Lambda(L)=\{0, 1.5\pm 1.323 i ,2\}$, i.e., $-L$ is marginally stable of corank $1$, and $\Lambda(L_s)=\{0, 0.7192,1.5, 2.7808\}$, i.e., $L_s$ is psd of corank $1$.
	\end{example}

	Observe that Theorem~\ref{thm:signed-digraph1} and Corollary~\ref{cor:weight-bal1} do not explicitly assume that $\mathcal{G}$ is strongly connected. 
	As already mentioned, it follows from Property~2 in Proposition~\ref{prop:signed-prop} that for signed graphs strong connectivity of $\mathcal{G} $ (and irreducibility of $L$) does not always lead to ${\rm corank}(L)= 1$, see Example~\ref{ex:corank2}.
Similarly, ${\rm corank}(L)= 1$ by itself need not imply $L$ irreducible, even though it implies the existence of a rooted spanning tree. 
Irreducibility follows if in addition $L$ is weight balanced. 	
In particular, irreducibility of $L$ (and strong connectivity of $\mathcal{G}$) are implied by any of the conditions (i) or (ii) of Corollary~\ref{cor:weight-bal1}.
These properties are formalized in the following lemma.

\begin{lemma}\label{lemma:irreducible} 
Let $\mathcal{G}$ be a signed digraph with Laplacian $L$. 
\begin{enumerate}
    \item If $L$ is of corank $1$, then $\mathcal{G}$ has a rooted spanning tree.
    \item If $-L$ is EEP or if $L$ is weight balanced and of corank $1$, then $L$ is irreducible (and $ \mathcal{G} $ strongly connected).
\end{enumerate}		
\end{lemma}
The proof is in Appendix~\ref{app:TI}.	

	

	\subsection{Discrete-time protocol}
	\label{sec:discrete}
	
	In this section we use the notion of ES to establish equivalent results for the discrete-time system \eqref{eq:discrete_sys} on a signed graph $ \mathcal{G}$.
	We start by observing that also in this case, irreducibility of $W$ does not guarantee simplicity of the dominant eigenvalue, which in turn may hamper consensus even though $W$ is still marginally Schur stable.
	
	\begin{example}
		\label{ex:Wcorank2}
		Consider $ W = I- L/10$, where $L$ is given in Example~\ref{ex:corank2}. By construction $ W \mathbf{1} = W^\top \mathbf{1} = \mathbf{1}$ and $ W$ irreducible. 
		Nevertheless, $ \Lambda(W) =\{ 0.6, \, 0.6, \, 1, \, 1 \}$, i.e., the dominant eigenvalue has multiplicity 2 and $ W$ is marginally Schur stable.
		$\hfill\square$
	\end{example}

	The following lemma states that consensus is equivalent to marginal Schur stability of $W$ plus simplicity and strict dominance of the $ \lambda(W) =1 $ eigenvalue. 
	
	\begin{lemma}\label{le:nec:DT}
		Consider the system \eqref{eq:discrete_sys} and let the signed weighted adjacency matrix $W$ satisfy $W\mathbf{1}=\mathbf{1}$. Consensus is achieved if and only if $W$ is marginally Schur stable with the eigenvalue $\lambda(W) = 1$ which is simple and strictly dominant.
	\end{lemma}
The proof is in Appendix~\ref{app:TI}.

	\begin{example}
		\label{W:corank-nostrictdom}
		Consider $ W = I- L/2$, where $L$ is given in Example~\ref{ex:corank2}. By construction $ W \mathbf{1} = W^\top \mathbf{1} = \mathbf{1}$. $ \Lambda(W) =\{ -1, \, -1, \, 1, \, 1 \}$, i.e., $ \rho(W) = |\lambda_i(W)| $ for all $ i$. $W$ is marginally stable, but simplicity and strict dominance are missing. 
		Since $ \lim_{t\to\infty} W^t $ does not exist, the system \eqref{eq:discrete_sys} does not converge to consensus. 
		If instead we consider $ W$ of Example~\ref{ex:Wcorank2}, we have strict dominance of $ \lambda(W) =1 $, but with multiplicity 2. Also here no consensus is achieved for the system \eqref{eq:discrete_sys}.
		$\hfill\square$
	\end{example}

	\medskip
	
	Apart from consensus, when $ W\geq 0 $, a closely related version of \eqref{eq:discrete_sys} is commonly used to describe transition probabilities in Markov chains.
	Clearly when $ W$ is not nonnegative then any probabilistic interpretation associated to $W$ is lost. 
	However, if $ W^t >0$ for $ t \geq t_o $, any sufficiently long downsampling of the system \eqref{eq:discrete_sys} can still be considered a well-posed transition matrix, provided $W$ is eventually stochastic.

	Neither the undirected graph case, nor the digraph case have been analyzed in the literature so far, so we discuss both in the following.
	
	\medskip
	
	\subsubsection{Signed undirected graph case}
	
	The following necessary and sufficient condition is the analogous of Theorem~\ref{thm:signed-undirected-cont} for discrete-time systems.
	\begin{theorem} 
	\label{thm:symm-sign}
		Consider an undirected signed graph $ \mathcal{G}$ of signed adjacency matrix $W$ which is symmetric, 
		and such that $ W \bfone =\bfone$. 
		$W$ is marginally Schur stable with $ \lambda(W)=1 $ simple and strictly dominant iff $W$ is EDS.
	\end{theorem}
The proof is in Appendix~\ref{app:TI}.


	\medskip
	
	\subsubsection{Signed digraph case}

	As for Laplacians, on signed digraphs the sufficient condition for consensus that can be obtained from ES matrices is in general not a necessary condition.

	\begin{theorem}
		\label{eq:ev-stoch-matrix}
		Consider a signed digraph $ \mathcal{G}$ of adjacency matrix $W$ such that $ W \mathbf{1} = \mathbf{1}$.
		If $ W$ is ES, then it is marginally Schur stable with $ \lambda(W)=1 $ simple and strictly dominant, and the system \eqref{eq:discrete_sys} converges to consensus.
		Viceversa, if $W$ is marginally Schur stable with $ \lambda(W)=1 $ simple and strictly dominant, then $ W \in \mathcal{PF}$. 
	\end{theorem}
The proof is in Appendix~\ref{app:TI}.

	The following example shows that the condition of eventual stochasticity in Theorem~\ref{eq:ev-stoch-matrix} is not necessary for marginal stability with $ \lambda(W)=1 $ simple and strictly dominant.
	\begin{example}
		The matrix
		\[
		W=\begin{bmatrix}
			0.0893   & 0.8036 &  -0.2679  &  0.375 \\
			0.9932  &  0.0068     &    0    &     0 \\
			0.8625  & -0.0375  &  0.1   & 0.075 \\
			0.114  &  0.8547   &      0  &  0.0313 \\
		\end{bmatrix}
		\]
		is such that $ W\bfone = \bfone$. Since $ \Lambda (W) =\{ -0.4603 \pm 0.3218i, \, 
		0.148, \, 1\}$, $W$ is marginally Schur stable with $ \lambda(W)=1 $ simple and strictly dominant. Furthermore, $ W \in \mathcal{PF}$, but $ W^\top  \notin \mathcal{PF}$, hence $W$ is not an ES matrix.
		$\hfill\square$
	\end{example}

	The following example shows instead that the weaker condition $ W\in \mathcal{PF} $ does not suffice for marginal stability.

	\begin{example}
		The matrix
		\[
		W=\begin{bmatrix}
			0.0368    &     0    &  0.659   &   0.3042 \\
			0    &  0.0872    &  3.1947   &  -2.2819 \\
			0   &   0.9895   &   0.0105     &      0 \\
			0.9322      &     0     &      0   &   0.0678 \\
		\end{bmatrix}
		\]
		is such that $ W\bfone = \bfone$, $ \Lambda (W) =\{-1.5494, \, -0.8928, \, 1, \, 1.6447\}$, meaning that $ \rho(W)  =1.6447$ of right eigenvector 
		$
		\bm{\nu} = \begin{bmatrix} 
			0.2303 & 
			0.8242 & 
			0.4991 &
			0.1362
		\end{bmatrix}^\top  >0.
		$
		Therefore $ W \in \mathcal{PF}$, but $W$ unstable. Notice that $W$ has two positive right eigenvectors and obviously $ W^\top  \notin \mathcal{PF}$.
		$\hfill\square$
	\end{example}

	Also for signed stochastic matrices the weight balanced case (here $ W\bfone = W^\top  \bfone = \bfone $) leads instead to a necessary and sufficient condition.
	
	\begin{corollary}
		\label{cor:weight-bal-DT} 
		Consider a signed digraph $ \mathcal{G}(W)$ such that $W$ is weight balanced and $W\bfone =\bfone$.
		Then the following conditions are equivalent:
		\begin{enumerate}[(i)]
			\item $W$ is EDS;
			\item $W$ is marginally Schur stable with $ \lambda(W)=1 $ simple and strictly dominant;
		\end{enumerate}
		Furthermore, if $W$ is normal, then (i) and (ii) are equivalent to:
		\begin{enumerate}[(i)]
			\item [(iii)]$ I - W^\top W $ is psd of corank 1. 
		\end{enumerate}
	\end{corollary}
The proof is in Appendix~\ref{app:TI}.

	\begin{lemma}
		\label{lem:irreduc_W_suff}
		Consider a signed digraph $ \mathcal{G} $ of adjacency matrix $ W$ such that $ W \mathbf{1} = \mathbf{1} $. 
		If $W$ is ES or if $W$ is weight balanced with $ \lambda(W) =1 $ which is simple and strictly dominant, then $W$ is irreducible (and $ \mathcal{G} $ strongly connected).
	\end{lemma}
	Since $ W$ ES implies $ W$ EP, the proof resembles that of Lemma~\ref{lemma:irreducible} and is therefore omitted. 
	

	\section{Consensus on time-varying signed digraphs}
	\label{sec:TV}
	
	\subsection{Problem formulation}
	
	Given an agent set $\mathcal{V}=[n]$, we consider a group of signed digraphs $\mathbb{G}=\{\mathcal{G}_1,\mathcal{G}_2,\dots,\mathcal{G}_{m}\}$ over $\mathcal{V}$. We consider two consensus protocols, one in continuous-time and one in discrete-time, in which the time-varying system is approximated as a switching of the underlying graph over $\mathbb{G}$. 
	
	\begin{enumerate}[1.]
		\item \textbf{Continuous-time protocol}. For each graph $\mathcal{G}_k$, the associated signed Laplacian matrix $L_k$ is defined as in \eqref{eq:Laplacian1}.
		The corresponding set of signed Laplacian matrices is denoted $\mathbb{L}= \{L_1,L_2,\dots,L_m\}$. 
		The system can be written as
		\begin{equation}\label{eq:form_c}
			\dot{\mathbf{x}}=-L_{\sigma(t)}\mathbf{x},\quad \mathbf{x}\in\mathbb{R}^{n},
		\end{equation}
		where $\sigma(\cdot)$ is a switching signal, which is a piecewise constant map, i.e., $\sigma:\mathbb{R}_{\geq}\mapsto [m]$. Piecewise constant means that any finite interval of $\mathbb{R}_\geq$ can have at most finitely many discontinuities, meaning that Zeno-like phenomena are avoided \cite{lin2009stability}. 
		
		\item \textbf{Discrete-time protocol}. The corresponding set of signed weighted adjacency matrices is denoted $\mathbb{W}=\{W_1,W_2,\dots,W_m\}$, where 
		\[W_k\mathbf{1}=\mathbf{1},\quad \forall k\in [m].\] 
		The discrete-time system is
		\begin{equation}\label{eq:form_d}
			\mathbf{x}(t+1)=W_{\sigma(t)}\mathbf{x}(t), \quad t\in\mathbb{Z}_{\geq}, \mathbf{x}(t)\in\mathbb{R}^{n},
		\end{equation}
		where $\sigma:\mathbb{Z}_{\geq}\mapsto [m]$ is also a switching signal with the same properties.
	\end{enumerate}

	For the time-varying systems \eqref{eq:form_c} and \eqref{eq:form_d}, in order to achieve consensus in the sense of Definition~\ref{def:cons} for an arbitrary switching signal $\sigma(\cdot)$, it is convenient to introduce the following notion of consensus set.
	
	\begin{definition}[Consensus set]
		The set $\mathbb{L}$ (resp. $ \mathbb{W}$) is said to be a consensus set for the system \eqref{eq:form_c} (resp. \eqref{eq:form_d}) if consensus is achieved for any arbitrary piecewise constant switching signal $\sigma(\cdot)$.
	\end{definition}
	
	\textbf{Problem of interest} for this section: find conditions under which the set $\mathbb{L}$ (resp. $ \mathbb{W}$) is a consensus set for the system \eqref{eq:form_c} (resp. \eqref{eq:form_d}).
	

	\subsection{Continuous-time protocol}\label{sec:CT}
	
	For the system \eqref{eq:form_c}, in order to achieve consensus under any switching signal, each subsystem $\dot{\mathbf{x}}=-L_k\mathbf{x}$ must achieve consensus. This observation leads to the following necessary condition, whose proof is a straightforward application of Lemma \ref{le:nec}.
	
	\begin{lemma}\label{tem:nec}
		If $\mathbb{L}$ is a consensus set for the system \eqref{eq:form_c}, then $-L_k$ is marginally stable of corank $1$ for all $k\in[m]$.
	\end{lemma}
	
	In the following we will focus on weight balanced digraphs, i.e., $L_k^\top  \mathbf{1}=0, k\in[m]$.
	As argued above, in order to make $\mathbb{L}$ a consensus set, each $L_k$ must be marginally stable of corank $1$. In addition, if $\mathcal{G}_k$ is weight balanced, by Corollary~\ref{cor:weight-bal1} and Lemma~\ref{lemma:irreducible}, the following corollary of Lemma~\ref{tem:nec} is obtained.
	
	\begin{corollary}
		Suppose that each $\mathcal{G}_k\in\mathbb{G}$ is weight balanced. If $\mathbb{L}$ is a consensus set, then for all $k\in[m]$, $-L_k$ is EEP, which implies that $\mathcal{G}_k$ is strongly connected.
	\end{corollary}
	
	However, only EEP of each $L_k$ is not enough. This can be seen from the following example.
	
	\begin{figure}[ht]
		\centering
		\includegraphics[trim=0cm 0cm 0cm 0cm,clip=true,width=9cm]{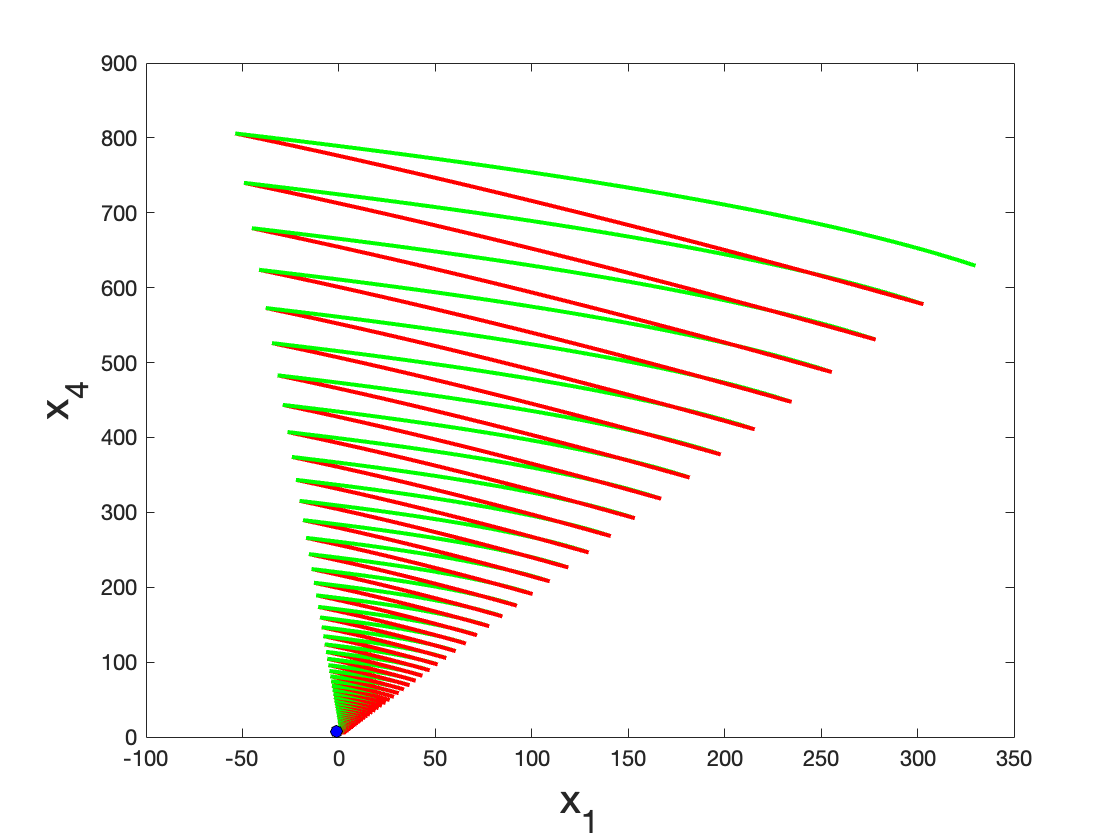}
		\caption{$2$-dimensional slice of a trajectory for Example. \ref{ex:ce1}. The dark dot represents the initial state. Line color changes every time the system switches.}
		\label{fig:1}
	\end{figure}
	
	\begin{example}\label{ex:ce1}
		Consider the system \eqref{eq:form_c} on $\mathbb{G}=\{\mathcal{G}_1,\mathcal{G}_2\}$, with the following Laplacians
		\begin{equation*}
			\begin{aligned}
				L_1 &=\left[ \begin{array}{cccc}
					0.23  & 0 & -0.28 & 0.05 \\
					-0.01 & 0.03 & 0.02 & -0.04 \\
					0.05 & -0.03 & 0.04 & -0.06 \\
					-0.27 & 0 & 0.22 & 0.05
				\end{array}\right],\\
				L_2 &=\left[ \begin{array}{cccc}
					0.96 & -0.27 & 0 & -0.69 \\
					-0.44 & 0.44 & 0 & 0 \\
					-0.43 & -0.17 & 0.07 & 0.53\\
					-0.09 & 0 & -0.07 & 0.16
				\end{array}\right].
			\end{aligned}
		\end{equation*}
		$L_1,L_2$ are both weight balanced. Moreover, $0$ is a simple rightmost eigenvalue of $-L_1,-L_2$, which by Corollary~\ref{cor:weight-bal1} means that $-L_1,-L_2$ are EEP. Let the switching signal be
		\[\sigma(t)=\left\{ \begin{array}{cc}
			1, & \text{if } t\in [2d,2d+1);\\
			2,    & \text{if } t\in [2d+1,2d+2),
		\end{array}\right. d=0,1,\dots \]
		Let the initial state be $\mathbf{x}(0)=[-1,2,-4,7]^\top $. Fig. \ref{fig:1} shows the trajectory of $(x_1(t),x_4(t))$. It is seen that the switching system diverges. 
		$\hfill\square$
	\end{example}
	
	One condition we can add to EEP is normality of the signed Laplacians. 
	
	\begin{remark}
		\label{rem:normal}
		Given a normal matrix, any real right eigenvector is also a left eigenvector corresponding to the same eigenvalue \cite{horn2012matrix}. Thus, any normal signed Laplacian is weight balanced.
	\end{remark} 
	The following theorem is the main result of this Section.
	
	\begin{theorem}\label{th:form_c}
		Consider a set of signed digraphs $\mathbb{G}=\{\mathcal{G}_1,\dots,\mathcal{G}_m\}$ with the corresponding Laplacians $\mathbb{L}=\{ L_1,\dots,L_m\} $. If $-L_k$ is a EEP, normal matrix for each $k\in[m]$, then $\mathbb{L}$ is a consensus set for the system \eqref{eq:form_c}.
	\end{theorem}
The proof is in Appendix~\ref{app:TV}.

	\begin{remark}\label{re:cons}
		Any symmetric matrix is normal. Therefore, if for all $k\in [m]$, $-L_k$ is symmetric and EEP, $\mathbb{L}$ is then a consensus set. 
	\end{remark}
	
	Note that the proof of Theorem \ref{th:form_c} is similar to that in \cite{olfati2004consensus}. The requirement that $L_k+L_k^\top ,k\in[m]$ is psd of corank $1$ is key to the proof. This condition is equivalent to the existence of a common quadratic Lyapunov function with matrix which is diagonal and equal to $I$. To satisfy this condition, the normality of each Laplacian is sufficient but not necessary, see Sect.~\ref{sec:CLF} for a counterexample.
	
	One may also ask if the condition $L_k+L_k^\top$ is psd of corank $1$ for all $ k\in[m]$, is a necessary condition. The answer is no, see the following example.
	
	\begin{example}
		Consider the system \eqref{eq:form_c} on $\mathbb{G}=\{\mathcal{G}_1,\mathcal{G}_2\}$ with the corresponding Laplacians $L_1$ given in Example \ref{ex:ce1} and $L_2=I-\frac{1}{4}\mathbf{1}\mathbf{1}^\top $. 
		It can be checked that $L_1+L_1^\top $ is not psd. For any given switching signal $\sigma(t)$, without loss of generality, suppose the signals switch in the sequence $1,2,1,2,\dots$ with switching time $0<t_1<t_2<t_3<\dots$. Let 
		\[t_1=s_1,t_2-t_1=\tau_1,t_3-t_2=s_2,t_4-t_3=\tau_2,\dots \]
		The solution can be explicitly written as
		\begin{equation*}
			\begin{aligned}
				&\mathbf{x}(t)= \\
				& \left\{\begin{array}{c}
					[(1-e^{-\sum\limits_{d=1}^\kappa \tau_d}) \frac{1}{n}\mathbf{1}\mathbf{1}^\top +e^{-\sum\limits_{d=1}^\kappa \tau_d} e^{-(s'+\sum\limits_{d=1}^\kappa s_d)L_1}]\mathbf{x}(0), \\
					\text{if } t=\sum\limits_{d=1}^{\kappa}(s_d+\tau_d)+s',\quad  0\leq s'<s_{\kappa+1}; \\
					
					[(1-e^{-\tau'-\sum\limits_{d=1}^\kappa \tau_d}) \frac{1}{n}\mathbf{1}\mathbf{1}^\top +e^{-\tau'-\sum\limits_{d=1}^\kappa \tau_d} e^{(-\sum\limits_{d=1}^{\kappa+1} s_d)L_1}]\mathbf{x}(0),\\
					\text{if } t=\sum\limits_{d=1}^{\kappa}(s_d+\tau_d)+s_{\kappa+1}+\tau', \quad 0\leq \tau'<\tau_{\kappa+1}.
				\end{array}\right.
			\end{aligned}
		\end{equation*}
		Then
		\begin{equation}\label{eq:limit}
			\lim_{t\to\infty}\mathbf{x}(t)=[(1-e^{-\sum\limits_{d=1}^\infty \tau_d})\frac{1}{n}\mathbf{1}\mathbf{1}^\top  +e^{-\sum\limits_{d=1}^\infty \tau_d}e^{-(\sum\limits_{d=1}^\infty s_d)L_1}]\mathbf{x}(0).
		\end{equation}
		It is easy to see that:
		\begin{itemize}
			\item if $\sum_{d=1}^{\infty}\tau_d=\infty$, the second part of \eqref{eq:limit} is $0$, which yields $\lim_{t\to\infty}\mathbf{x}(t)=\frac{1}{n}\mathbf{1}\mathbf{1}^\top \mathbf{x}(0)$; 
			\item if $\sum_{d=1}^{\infty}\tau_d\not=\infty$, it must be $\sum_{d=1}^{\infty}s_d=\infty$. We then have
			\[e^{-(\sum\limits_{d=1}^\infty s_d)L_1}=\frac{1}{n}\mathbf{1}\mathbf{1}^\top,\]
			since $-L_1$ is EEP. According to \eqref{eq:limit}, it holds $\lim_{t\to\infty}\mathbf{x}(t)=\frac{1}{n}\mathbf{1}\mathbf{1}^\top \mathbf{x}(0)$.
		\end{itemize}
		No matter in which case, it always holds that $\lim_{t\to\infty}\mathbf{x}(t)=\frac{1}{n}\mathbf{1}\mathbf{1}^\top \mathbf{x}(0)$. As $\sigma(t)$ is arbitrarily chosen, $\mathbb{L} =\{ L_1, \, L_2 \} $ is a consensus set.
		$\hfill\square$
	\end{example}

	
	\subsection{Discrete-time protocol}

	For the discrete-time system  \eqref{eq:form_d}, in order to achieve consensus under any switching signal, each subsystem  $\mathbf{x}(t+1)=W_k\mathbf{x}(t)$ must achieve consensus, which leads to the following necessary condition.
	
	\begin{lemma}\label{lem:nec:DT}
		Suppose $W_k\mathbf{1}=\mathbf{1}$ for all $k\in[m]$. If $\mathbb{W}$ is a consensus set for the system \eqref{eq:form_d}, then $W_k$ is marginally Schur stable with $ \lambda(W) =1 $ which is simple and strictly dominant for all $k\in[m]$.
	\end{lemma}
	The lemma is proven straightforwardly from Lemma~\ref{le:nec:DT}.
	From Theorem~\ref{eq:ev-stoch-matrix}, it implies that it must be  $W_k\in\mathcal{PF}_n$ $ \forall \, k \in [m]$. 
	In the weight balanced case, combining Lemma~\ref{lem:nec:DT} with Lemma~\ref{lem:irreduc_W_suff} a corollary follows.

	\begin{corollary}
		Suppose that each $\mathcal{G}_k\in\mathbb{G}$ is weight balanced. If $\mathbb{W}$ is a consensus set, then for all $k\in[m]$, $W_k$ is EDS and hence EP, which implies that $\mathcal{G}_k$ is strongly connected.
	\end{corollary}
	
	However, just like in the continuous-time case, only EP/EDS is still not enough, as the following example shows.
	
	\begin{example}\label{ex:cout_ex_d1}
		Consider the system \eqref{eq:form_d} on $\mathbb{G}=\{\mathcal{G}_1,\mathcal{G}_2\}$, with 
		\begin{equation*}
			\begin{aligned}
				W_1 &=\left[ \begin{array}{cccc}
					0.79 & 0 & 0.25 & -0.04 \\
					0.01 & 0.97 & -0.02 & 0.04 \\
					-0.04 & 0.03 & 0.95 & 0.06 \\
					0.24 & 0 & -0.18 & 0.94
				\end{array}\right],\\
				W_2 &=\left[ \begin{array}{cccc}
					0.42 & 0.14 & 0.02 & 0.42 \\
					0.23 & 0.68 & 0 & 0.09 \\
					0.28 & 0.17 & 0.92 & -0.37\\
					0.07 & 0.01 & 0.06 & 0.86
				\end{array}\right].
			\end{aligned}
		\end{equation*}
		The switching signal is $1,2,1,2,\dots$. It can be verified that $W_1\mathbf{1}=W_1^\top \mathbf{1}=\mathbf{1}$ and so is for $W_2$. Furthermore, $\rho(W_1)=1, \rho(W_2)=1$ and $1$ is a simple eigenvalue. Therefore, $W_1$ and $W_2$ are both EP. On the other hand, $\rho(W_2W_1)=1.1008>1$. Thus, the switching system \eqref{eq:form_d} diverges. 
		$\hfill\square$
	\end{example}
	
	
	\begin{theorem}\label{th:form_d}
		Consider a signed digraph set $\mathbb{G}=\{\mathcal{G}_1,\dots,\mathcal{G}_m\}$ with the corresponding weighted adjacency matrices $\mathbb{W} = \{ W_1,\dots,W_m\}$. If, for all $k\in[m]$, $W_k$ is a ES, normal matrix, then $\mathbb{W}$ is a consensus set for the system \eqref{eq:form_d}.
	\end{theorem}
The proof is in Appendix~\ref{app:TV}.

	\begin{remark}\label{re:disc}
		As in Remark \ref{re:cons}, if $W_k$ is symmetric and ES for all $k\in[m]$, $\mathbb{W}$ is a consensus set.
	\end{remark}
	
	According to the proof of Theorem \ref{th:form_d}, we can see that it is enough to have $I-W_k^\top  W_k$ psd and of corank $1$ for all $k\in[m]$. Therefore, the condition that each $W_k$ is normal is sufficient but not necessary, which is also shown in the following example.
	
	\begin{example}\label{ex:cout_ex_d2}
		Let $\mathbb{G}=\{\mathcal{G}_1,\mathcal{G}_2\}$ corresponding to $\mathbb{W}=\{W_1,W_2\}$, with 
		\begin{equation*}
			\begin{aligned}
				W_1=\left[\begin{array}{ccc}
					0.1  &  0.2 & 0.7\\
					0.3   & 0.5 & 0.2\\
					0.6 & 0.3 & 0.1
				\end{array}\right], \quad W_2=W_1^\top .
			\end{aligned}
		\end{equation*}
		$W_1$ and $W_2$ are positive and doubly stochastic. Obviously $\mathbb{W}$ is a consensus set for the system \eqref{eq:form_d} (since $\mathcal{W}$ is a finite set of stochastic matrices with positives columns \cite{wu2007synchronization}), but $W_1$ and $W_2$ are not normal. 
		$\hfill\square$
	\end{example}
	
	Again, also the condition that $I-W_k^\top  W_k$ is psd and of corank $1$ for all $k\in[m]$ is not necessary. 
	
	\begin{example}\label{ex:cout_ex_d3}
		Consider $\mathbb{G}=\{\mathcal{G}_1,\mathcal{G}_2\}$ with the corresponding signed weighted adjacency matrices $W_1$ given in Example \ref{ex:cout_ex_d1}, and $W_2=\frac{1}{4}\mathbf{11}^\top $. Obviously, once the switching signal becomes $2$, the system achieves consensus, no matter what the signal is at other times. On the other hand, if the signal never switches to $2$, it must be $\sigma(t)=1,t\geq 0$, for which the system also achieves consensus. Therefore, $\mathbb{W} =\{ W_1, \, W_2 \} $ is a consensus set. However, it can be verified that $\rho(W_1^\top  W_1)>1$, which means that $I-W_1^\top  W_1$ is not psd. 
		$\hfill\square$
	\end{example}

	\section{More general sufficient conditions for the time-varying case}
	\label{sec:CLF}
	
	It is well-known \cite{liberzon2003switching,lin2009stability} that a sufficient condition for a collection of Hurwitz (or Shur) stable linear systems to be stable as a switching system is to have a Common Lyapunov Function (CLF).
	As the proofs show, also the conditions of Theorems~\ref{th:form_c} and~\ref{th:form_d} are based on the existence of a  CFL, which is diagonal and with diagonal weights all equal to 1. While such sufficient condition is of interest per se, as it has an equivalent algebraic formulation in terms of EEP/ES normal matrices, it can be relaxed to any (non-diagonal) CLF, allowing us to relax the assumption of normality of the $ L_k $ and $ W_k $. In this way, the class of switching signed graphs for which time-varying consensus is achieved is enlarged significantly.
	Existence of a CLF can be expressed as feasibility of a systems of Linear Matrix Inequalities (LMI).
	
	\subsection{Continuous-time case}

	Denote $ Q \in \mathbb{R}^{n-1 \times n } $ a matrix whose rows form an orthonormal basis for $ {\rm span}(\mathbf{1})^\perp$, i.e.,
	\[QQ^\top=I_{n -1}, \quad Q\mathbf{1}=\mathbf{0}.\]
	
	\begin{theorem}\label{th:CLF1}
		Consider a set of signed digraphs $\mathbb{G}=\{\mathcal{G}_1,\dots,\mathcal{G}_m\}$ with the corresponding Laplacians $\mathbb{L}= \{ L_1,\dots,L_m\}$. Assume $-L_k$ is a weight balanced EEP matrix for each $k\in[m]$. If $ \exists $ $P =P^\top \succ 0 $ such that
		\beq
		-Q L_k P Q^\top -  Q P L_k^\top  Q^\top \prec 0 , \quad k =1, \ldots, m 
		\label{eq:CLF1}
		\eeq
		then $ V(\mathbf{x}) = \mathbf{x}^\top  P \mathbf{x} $ is a CLF and hence $\mathbb{L}$ is a consensus set for the system \eqref{eq:form_c}.
	\end{theorem}
The proof is in Appendix~\ref{app:TV}.

	\begin{example}
		
		None of the following two 
		weight-balanced  Laplacian matrices whose negation is EEP
		\[
		\begin{aligned}
			L_1& =\begin{bmatrix}
				0.3388  & -0.5673   &  0.2285 \\
				0    & 0.5673   & -0.5673 \\
				-0.3388     &     0   &  0.3388
			\end{bmatrix} \\
			L_2 & =\begin{bmatrix}
				0.3937  &  -0.6570   &  0.2633 \\
				-0.1862   &  0.6570  &  -0.4708 \\
				-0.2075     &     0  &   0.2075
			\end{bmatrix}
		\end{aligned}
		\]
		is normal.
		For them 
		\[
		P =\begin{bmatrix}
			72.7760  &   18.9053  &   -6.5237 \\
			18.9053  &   68.4564  &   -2.2041 \\
			-6.5237  &   -2.2041  &   93.8854
		\end{bmatrix}
		\]
		is a CLF.
		$\hfill\square$
	\end{example}
	
	Similarly to what happens to switching systems of Hurwitz matrices, existence of a quadratic CLF is sufficient but not necessary, and higher order homogeneous polynomial CLF can be used to relax the conservatism of quadratic design \cite{Chesi1}. Following \cite{Cla-HPCLF1}, to build a $ 2r$-th order homogeneous CLF we can make use of ($r$-times) Kronecker products  and Kroneker sums, defining
	\[
	\begin{aligned}
		\hat{\mathbf{x}} & = \mathbf{x} \otimes \ldots \otimes \mathbf{x} \\ 
		\hat L_k & = L_k \oplus \ldots \oplus L_k  \\
		\hat Q & = Q \otimes \ldots \otimes Q .
	\end{aligned}
	\]
	Easy calculations then give the following extension of Theorem~\ref{th:CLF1}.
	\begin{corollary}\label{cor:CLF2}
		Consider a set of signed digraphs $\mathbb{G}=\{\mathcal{G}_1,\dots,\mathcal{G}_m\}$ with the corresponding Laplacians $\mathbb{L}= \{ L_1,\dots,L_m\}$. Assume $-L_k$ is a weight balanced EEP  matrix for each $k\in[m]$. If $ \exists $ $ \hat P \in \mathbb{R}^{n^r \times n^r}$, $\hat P=\hat P^\top  \succ 0$ such that
		\[
		- \hat Q \hat L_k \hat P \hat Q^\top  - \hat Q \hat P \hat L_k^\top  \hat Q^\top  \prec 0 , \quad k =1, \ldots, m 
		\]
		then $ V(\mathbf{x}) = {\hat{\mathbf{x}}}^\top  \hat P \hat{\mathbf{x}} $ is a CLF and hence $\mathbb{L}$ is a consensus set for the system \eqref{eq:form_c}.
	\end{corollary}

	\begin{example}
		The following two signed Laplacian matrices
		\[
		\begin{aligned}
			L_1& =\begin{bmatrix}
				0.6667  &  0.2440  & -0.9107 \\
				-0.9107  &  0.6667  &  0.2440 \\
				0.2440 &  -0.9107  &  0.6667
			\end{bmatrix} \\
			L_2 & =\begin{bmatrix}
				2.6111  &  0.4746  & -3.0857 \\
				-3.0857  & -0.3056  &  3.3913 \\
				0.4746  & -0.1691  & -0.3056
			\end{bmatrix}
		\end{aligned}
		\]
		do not admit a quadratic CLF, but they admit a CLF which is a homogeneous polynomial of order 4. In fact their projections $ \bar L_k = Q L_k Q^\top $ correspond to a famous example of 2D switching system not admitting a quadratic CLF \cite{Dayawansa1}:
		\[
		\bar L_1 =\begin{bmatrix} 1 &1  \\ -1 &1\end{bmatrix} , \qquad \bar L_2 =\begin{bmatrix} 1 &6  \\ -1/6& 1\end{bmatrix} 
		\]
		$\hfill\square$
	\end{example}

	\subsection{Discrete-time case}
	
	Also for the discrete-time case the consensus set stability can be extended beyond normality via CLF.
	
	\begin{theorem}\label{th:CLF-DT1}
		Consider a set of signed digraphs $\mathbb{G}=\{\mathcal{G}_1,\dots,\mathcal{G}_m\}$ with the corresponding EDS matrices $\mathbb{W} =\{ W_1,\dots,W_m\}$. If $ \exists $ $P =P^\top  \succ 0$ such that
		\beq
		Q W_k^\top  P W_k Q^\top  - Q P Q^\top   \prec  0, \quad k = 1, \ldots, m 
		\label{eq:CLF-DT1}
		\eeq
		then $ V(\mathbf{x}) = \mathbf{x}^\top  P \mathbf{x} $ is a CLF and hence $\mathbb{W}$ is a consensus set for the system \eqref{eq:form_d}.
	\end{theorem}
The proof is in Appendix~\ref{app:TV}.

	Denoting 
	\[
	\hat W_k  = W_k \otimes \ldots \otimes W_k, 
	\]
	we obtain also the discrete-time equivalent of Corollary~\ref{cor:CLF2}.

	\begin{corollary}\label{cor:CLF-DT2}
		Consider a set of signed digraphs $\mathbb{G}=\{\mathcal{G}_1,\dots,\mathcal{G}_m\}$ with the corresponding EDS matrices $\mathbb{W}= \{ W_1,\dots,W_m\} $.  If $ \exists $ $ \hat P \in \mathbb{R}^{n^r \times n^r}$, $\hat P=\hat P^\top  \succ 0$ such that
		\[
		\hat Q \hat W_k^\top   \hat P \hat W_k \hat Q^\top  - \hat Q \hat P  \hat Q^\top  \prec 0 , \quad k =1, \ldots, m 
		\]
		then $ V(\mathbf{x}) = {\hat{\mathbf{x}}}^\top  \hat P \hat{\mathbf{x}} $ is a CLF and hence $\mathbb{W}$ is a consensus set for the system \eqref{eq:form_d}.
	\end{corollary}

	\begin{example}
		The following two signed EDS matrices
		\[
		\begin{aligned}
			W_1& =\begin{bmatrix}
				0.2000  &   0.1691  &  0.6309 \\
				0.6309   &  0.2000  &   0.1691 \\
				0.1691  &   0.6309   &  0.2000 
			\end{bmatrix} \\
			W_2 & =\begin{bmatrix}
				-0.5778   &  0.0768   &  1.5010 \\
				1.5010   &  0.5889   & -1.0898 \\
				0.0768   &  0.3343   &  0.5889
			\end{bmatrix}
		\end{aligned}
		\]
		do not admit a quadratic CLF, but they admit a CLF which is a homogeneous polynomial of order 4. In this case their projections $ \bar W_k = Q W_k Q^\top $ correspond to another known example of uniformly stable switching system not admitting a quadratic CLF \cite{Theys2005}:
		\[
		\bar W_1 =\begin{bmatrix} -0.2 &-0.4  \\ 0.4 &-0.2\end{bmatrix} , \qquad \bar W_2 =\begin{bmatrix} -0.2 & -2.4  \\ 1/15& -0.2\end{bmatrix}.
		\]
		$\hfill\square$
	\end{example} 
	
	\section{Extension to bipartite consensus}
	\label{sec:bipartite}

	Consider a signed digraph $ \mathcal{G}$.
	Given a diagonal signature matrix $ S = {\rm diag}(\mathbf{s} ) $, consider the following generalizations of $L$ and $W$: 
	\begin{equation}
		\label{eq:Lb}
		L_b = S L S \quad \text{and} \quad 
		W_b = S W S ,
	\end{equation}
	and the corresponding generalizations of \eqref{eq:ode:laplacian} or \eqref{eq:discrete_sys}:
	\begin{equation}
		\label{eq:ode:laplacian-b}
		\dot{\mathbf{x}} = - L_b \mathbf{x} 
	\end{equation}
	and
	\begin{equation}
		\label{eq:discrete_sys-b}
		\mathbf{x}(t+1) = W_b \mathbf{x} (t) .
	\end{equation}
	Notice that $L_b $ does not obey to the rule \eqref{eq:Laplacian1}. In fact, $L_b = S L S = \Sigma - S AS $, and in general $  \sigma^{\rm in}_i = \sum_{j=1}^n A_{ij} \neq \sum_{j=1}^n [\mathbf{s}]_i [\mathbf{s}]_j A_{ij}$. 
	In matrix form, this read $ L_b \mathbf{s} =0$. 
	Similarly, in place of $ W \mathbf{1} = \mathbf{1} $, we now have $ W_b \mathbf{s} = \mathbf{s}$.
	
	The systems \eqref{eq:ode:laplacian-b} or \eqref{eq:discrete_sys-b} are associated to a more general form of consensus, called bipartite consensus  \cite{Altafini2013Consensus}.
	
	\begin{definition}
		\label{def:cbip-ons}
		We say that the system \eqref{eq:ode:laplacian-b} or \eqref{eq:discrete_sys-b} achieves bipartite consensus if for all $\mathbf{x}(0)\in\mathbb{R}^n$, there exist $\alpha\in\mathbb{R}$ and a diagonal signature matrix $ S = {\rm diag}(\mathbf{s}) $ such that $\mathbf{x}^\ast = \lim_{t\to\infty}\mathbf{x}(t)=\alpha \mathbf{s}$.
	\end{definition}
	Bipartite consensus corresponds to all agents achieving the same value in absolute value but possibly with different sign: $[\mathbf{x}^\ast]_i = \pm \alpha$, see \cite{Altafini2013Consensus}.
	In particular, each signature vector $ \mathbf{s}$ determines a ``bipartition class'' i.e., a splitting of the $n$ agents into two disjoint subgroups (up to a global symmetry $ \mathbf{s} \to - \mathbf{s}$).

	Several properties of $ L_b $ and $ W_b $ follow straightforwardly from \eqref{eq:Lb} and Proposition~\ref{prop:signed-prop}.
	For instance, for $L_b $ we have:
	
	\begin{proposition}
	\label{prop:bip}
		Given a signed digraph $ \mathcal{G}$, of Laplacian $L$, consider the associated matrix $ L_b $ in \eqref{eq:Lb} corresponding to $ S = {\rm diag}(\mathbf{s})$ and the system \eqref{eq:ode:laplacian-b}.
		Then we have:
		\begin{enumerate}
			\item $0$ is always an eigenvalue of $ L_b $ of right eigenvector $ \mathbf{s}$;
			\item $ \mathbf{x}(t) $ converges to $ \alpha \mathbf{s} $ iff $ \mathbf{z}(t) = S \mathbf{x}(t) $ converges to $ \alpha \mathbf{1}$, i.e., bipartite consensus is achieved by $ \mathbf{x}(t) $ iff consensus is achieved by $ \mathbf{z}(t)$.
			\item bipartite consensus is achieved iff $ -L_b $ is marginally stable of corank 1.
		\end{enumerate}
	\end{proposition}
The proof is in Appendix~\ref{app:TV}.

	All sufficient conditions for consensus given in the previous sections can be extended to bipartite consensus, provided we replace the notions of EP, EEP and ES with SEP, SEEP and SES.
	For instance the equivalent of Theorem~\ref{thm:signed-digraph1} and Corollary~\ref{cor:weight-bal1} is the following.
	
	\begin{theorem}
		Consider a signed digraph $ \mathcal{G}$ and a diagonal signature matrix $ S = {\rm diag}(\mathbf{s})$. 
		\begin{itemize}
			\item If $ - L_b $ is SEEP (w.r.t. $S$) then $ -L_b $ is marginally stable of corank 1;
			Viceversa, if $ -L_b $ is marginally stable of corank 1 then $ \exists \, d \geq 0 $ such that $ dI - L_b \in \mathcal{SPF}_n$.
			\item If, in addition, $ L_b \mathbf{s} = L_b^\top \mathbf{s} = 0 $, then the following two conditions are equivalent:
			\begin{enumerate}[(i)]
				\item $ -L_b $ is SEEP (w.r.t. $S$);
				\item $ -L_b $ is marginally stable of corank 1;
			\end{enumerate}
			\item If, in addition, $L_b $ is normal, then (i) and (ii) above are equivalent to 
			\begin{enumerate}
				\item[(iii)] $ (L_b+L_b^\top)/2 $ is psd of corank 1.
			\end{enumerate}
		\end{itemize}
	\end{theorem}
	The proof follows straightforwardly once the change of basis $ \mathbf{z}(t) = S \mathbf{x}(t) $ is performed. 
	
	\begin{remark}
		Recall that a signed digraph is said structurally balanced if all its directed cycles are positive (i.e., have an even number of negative edges). As shown in \cite{Altafini2013Consensus}, for the ``opposing'' signed Laplacian structural balance of the digraph is a necessary and sufficient condition for bipartite consensus. 
		The situation is very different for the ``repelling'' signed Laplacian $L$ and its generalization $ L_b $ we are considering in this paper.
		$ L $ can attain consensus even in presence of negative cycles, and it never achieves bipartite consensus. $ L_b $ can attain bipartite consensus even in presence of negative cycles, but it can also do so when all cycles are positive (in that case the corresponding $L$ is nonnegative). 
		Cycle sign is irrelevant for the ``repelling'' Laplacian $L$ and its generalization $ L_b$; the only thing that matters is the signature of the right PF eigenvector $ \mathbf{s}$.
	\end{remark}
	
	Also the time-varying case can be treated analogously, provided that all switching systems correspond to the same signature matrix $ S = {\rm diag}(\mathbf{s})$. 
	If we have the collection of signed digraphs $ \mathbb{G} = \{ \mathcal{G}_1 , \ldots, \mathcal{G}_m \}$ of Laplacians $ \mathbb{L}= \{ L_1, \ldots, L_m \}$, then consider the generalizations $ \mathbb{L}_b = \{  L_{b,1} , \ldots,  L_{b,m}  \}$, $ L_{b,k} = S L_k S$, $ k \in [m]$, and the system
	\begin{equation}\label{eq:form_c-b}
		\dot{\mathbf{x}}=-L_{b,\sigma(t)}\mathbf{x},\quad \mathbf{x}\in\mathbb{R}^{n},
	\end{equation}
	where $\sigma:\mathbb{R}_{\geq}\mapsto [m]$ is a piecewise constant switching signal.
	\begin{definition}[Bipartite consensus set]
		The set $\mathbb{L}_b$ is said to be a bipartite consensus set for the system \eqref{eq:form_c-b} w.r.t. the signature $ \mathbf{s}$ if bipartite consensus with node partition given by $ \mathbf{s}$ is achieved for any arbitrary piecewise constant switching signal $\sigma(\cdot)$.
	\end{definition}

	\begin{proposition}
		The set $ \mathbb{L}_b $ is a bipartite consensus set for the system \eqref{eq:ode:laplacian-b} w.r.t. the signature $ \mathbf{s}$ iff the set $ \mathbb{L} $ is a consensus set for the system \eqref{eq:ode:laplacian}.
	\end{proposition}
	Again the proof is immediate with the change of basis $ \mathbf{z}(t) = S \mathbf{x}(t) $.
	All conditions of Sections~\ref{sec:CLF} and~\ref{sec:bipartite} can then be translated into equivalent conditions for bipartite consensus for $ \mathbb{L}_b$.

	A similar reasoning applies also to the discrete-time case.

	\section{Conclusions}
	The conditions provided in this paper for checking the stability of signed (``repelling'') Laplacians are more general than those available in the literature, as they hold also for digraphs, and for time-varying signed Laplacians. They  are also insightful into the structure of these Laplacians, as they highlight the role of the dominant eigenpair of $L$.
	As computing dominant eigenpairs (through e.g. a power algorithm) is computationally cheaper than computing the entire spectrum of $L$, the tests provide also a computational advantage for large scale systems.

	Also in the time-varying case the behavior of signed Laplacians is qualitatively different from that of its standard ``nonnegative'' counterpart, in the sense that instabilities may arise due to the time dependence, making it even more relevant to seek some form of uniform convergence when dealing with this class of time-varying consensus problems.

	\appendices

\medskip
	
\appendices
\section{Time-invariant digraphs}
\label{app:TI}

	\prooftx{Proposition~\ref{prop:signed-prop}}
	Property 1 is true by construction. 
	Properties 2-5 can be shown through counterexamples (Examples~\ref{ex:corank2}-\ref{example-PF-notPF_unstable}).
	$\hfill\square$	
\medskip

	\prooftx{Lemma~\ref{le:nec}}
	We first prove the necessity. Notice that $0$ is an eigenvalue of $L$, with $\mathbf{1}$ included in its corresponding eigenvector space. The system $\dot{\mathbf{x}}=-L\mathbf{x}$ has the explicit solution $\mathbf{x}(t)=e^{-Lt}\mathbf{x}(0)$. If consensus is achieved, it must be $\lim_{t\to\infty}e^{-Lt}=\mathbf{1}\mathbf{c}^\top $ for some $\mathbf{c}\in\mathbb{R}^n$. Obviously $-L$ can not have eigenvalues with positive real parts, otherwise $e^{-Lt}$ will diverge. To prove marginal stability of $-L$, we use a contradiction argument. 
	First, assume that $0$ is not a simple root of the minimal polynomial of $-L$. Without loss of generalization, assume that $L$ can be decomposed to $L=CJC^{-1}$, where $J={\rm diag}(\begin{bmatrix} J_1 & J_2 & \dots & J_s \end{bmatrix} ) $ is the Jordan canonical form of $L$, with blocks $J_1=\left[\begin{array}{cc}
		0   &  1\\
		0  &  0
	\end{array}\right]$ and $J_2,\dots,J_s $ corresponding to eigenvalues of $L$ with positive real part.
	It then holds $e^{-Lt}=Ce^{-Jt}C^{-1}$, with $e^{-Jt}={\rm diag}(\begin{bmatrix}  e^{-J_1t}& e^{-J_2t}&\dots & e^{-J_st}  \end{bmatrix} )$. By straightforward calculations, we obtain 
	\begin{equation*}
		e^{-J_1t}=\left[\begin{array}{cc}
			1     & -t \\
			0     & 1
		\end{array}\right]
	\end{equation*}
	and
	\begin{equation*}
		\lim_{t\to\infty}{\rm diag}(\begin{bmatrix} ( e^{-J_2t}&\dots & e^{-J_st}  \end{bmatrix} )=\mathbf{0}.
	\end{equation*}
	Therefore,
	\[\lim_{t\to\infty} e^{-Lt}=\lim_{t\to\infty}C[e^{-Jt}]C^{-1}=\mathbf{1c}^\top  \]
	is impossible in presence of the term $e^{-J_1t}$.
	We then get the contradiction. Second, suppose that $di\in\Lambda(-L)$ for some $d\in\mathbb{R}$, where $i$ is the imaginary unit. Without loss of generality, the Jordan canonical form of $L$ then has a block $J_1$ such that $J_1=\left[\begin{array}{cc}
		di     & 1 \\
		0     & di
	\end{array}\right]$ or $J_1=\left[\begin{array}{cc}
		di     & 0 \\
		0     & -di
	\end{array}\right]$. Therefore,
	\begin{equation*}
		\lim_{t\to\infty} e^{-J_1t} = \left[\begin{array}{cc}
			e^{-dti}     & \ast \\
			0     & e^{-dti}
		\end{array}\right] \quad \text{or} \quad \left[\begin{array}{cc}
			e^{-dti}     & 0 \\
			0     & e^{dti}
		\end{array}\right],
	\end{equation*}
	where $\ast$ can be infinity. Again, no matter what case, it can not be $\lim_{t\to\infty} e^{-Lt}=\mathbf{1c}^\top  $, leading again to a contradiction. Therefore, $-L$ is marginally stable.
	To prove in addition that $L$ is of corank $1$, we  use another contradiction argument. Suppose that $-L$ is marginally stable of corank greater than or equal to $2$. 
	Let $J = CJC^{-1}$, where $J={\rm diag}(\begin{bmatrix} ( J_1 &\dots & J_{\mathrm{corank}(L)} & J_{\mathrm{corank}(L)+1} & \dots & J_s \end{bmatrix} )$ is the Jordan canonical form of $L$, with $J_\ell=[0]$ for all $\ell=1,\dots,\mathrm{corank}(L)$, and $J_\ell$, $\ell>\mathrm{corank}(L)$, corresponding to eigenvalues of $L$ with positive real parts. Then,
	\begin{equation}\label{eq:lim}
		\lim_{t\to\infty}e^{-L t} = \sum_{\ell=1}^{\mathrm{corank}(L)} \bm{\phi}_{\ell}\bm{\xi}_{\ell}^\top \ne \mathbf{1c}^\top ,
	\end{equation} 
	where $\bm{\phi}_{\ell},\bm{\xi}_{\ell}^\top $ are respectively the columns and rows of $C$ and $C^{-1}$. 
	The contradiction is then obtained and the necessity is proved. The sufficiency obviously follows from \eqref{eq:lim}.
	$\hfill\square$

\medskip	
	
	\prooftx{Theorem~\ref{thm:signed-digraph1}}
	From Lemma~\ref{lemma:ev_exp_pos}, $-L $ EEP means that 
	$ \exists $ a scalar $ d\geq 0 $ such that  $B = dI - L\stackrel{\vee}{>} 0 $. From $ L\bfone=0$, it is $ B\bfone = d \bfone - L \bfone = d\bfone $, i.e., $ \bfone $ is a right eigenvector of $B$ of eigenvalue $d$. 
	If $B\stackrel{\vee}{>} 0$ it means that $ \rho(B)$ is a simple eigenvalue of strictly positive left and right eigenvectors $ {\bf v}_r $ and $ {\bf v}_\ell$.
	From Lemma~\ref{lemma:no_two_pos-eig}, it must necessarily be $ {\bf v}_r = \alpha \bfone $ for some scalar $ \alpha$, and therefore $ \rho(B)=d$.
	Since $ \rho(B) $ is simple, all other eigenvalues of $L$ must have strictly positive real part.  The value of $ \mathbf{x}^\ast $ follows then straightforwardly.
	As for the second implication, if $- L$ is marginally stable and of corank 1, then $ 0 = \lambda_1(L) < {\rm Re}[\lambda_i(L)]$, $ i=2, \ldots, n$, and $ \bfone $ is the eigenvector relative to $0$.
	But then, provided $ d > \max_{i=2, \ldots, n} \frac{|\lambda_i(L)|^2}{2\real{\lambda_i(L)}}$,
	$ B = dI - L$ has $ \rho(B)=d$ of eigenvector $ \bfone$, hence $ B \in \mathcal{PF}$.
	$\hfill\square$

\medskip

	\prooftx{Corollary~\ref{cor:weight-bal1}}
	(i) $\Longrightarrow$ (ii): This direction is implied by Theorem~\ref{thm:signed-digraph1}. \\
	(ii) $\Longrightarrow$ (i): If $ -L $ marginally stable of corank 1, it follows from the proof of Theorem~\ref{thm:signed-digraph1} that for $ d > \max_{i=2, \ldots, n} \frac{|\lambda_i(L)|^2}{2\real{\lambda_i(L)}} >0 $,
	$ B= dI - L $ has $ \rho(B) =d $ which is a simple eigenvalue of right eigenvector $ \bfone$. 
	L weight balanced implies $ L \bfone = L^\top  \bfone =0$, which means that the argument can be repeated also for $L^\top  $ leading to $ B, \, B^\top  \in \mathcal{PF}$, i.e., from Theorem~\ref{thm:PFforEvPos}, $ B \epos 0 $ or, from Lemma~\ref{lemma:ev_exp_pos}, $ -L $ is EEP.
	%
	%
	
	\noindent (ii) $\Longleftrightarrow$ (iii): If $L$ is normal, then there exists an orthonormal matrix $ U$ such that $ L = U D U^\top  $, where, if  $ \mu_1, \ldots, \mu_\ell $ are the real eigenvalues of $ L $ and $ \nu_1\pm i \omega_1, \ldots , \nu_{\frac{n-\ell}{2}}  \pm i \omega_{\frac{n-\ell}{2}} $ are its complex conjugate eigenvalues:
	\begin{equation}
		D =\begin{bmatrix}  \mu_1 & \\ & \! \! \ddots & \\ & & \! \! \mu_\ell \\ 
			& & & \! \! \nu_1 & \omega_1 & \\
			& & & \! \! -\omega_1 & \nu_1 & \\
			& & & & & \ddots \\
			& & & & & & \! \! \nu_{\frac{n-\ell}{2}} & \omega_{\frac{n-\ell}{2}}  \\
			& & & & & & \! \! -\omega_{\frac{n-\ell}{2}} & \nu_{\frac{n-\ell}{2}} 
		\end{bmatrix}
		\label{eq:D-normal-diag}
	\end{equation}
	If follows that $ L_s = \frac{1}{2} (L + L^\top ) =  \frac{1}{2} U (D + D^\top ) U^\top  $ and therefore that $ {\rm Re}[\lambda_i (L) ] = \lambda_i (L_s) $. 
	$\hfill\square$

\medskip
\prooftx{Corollary~\ref{lem:psd-EP}}
First, observe that, since $L_s$ is psd, then $\mathbf{x}^\top L_s \mathbf{x} = 0$ iff $L_s \mathbf{x}=0$ 
\cite[Observation 7.1.6.]{HornJohnson2013}. Moreover, since $L_s$ is of corank $1$ and $L \mathbf{1} =0$ (which implies $\mathbf{1}^\top L_s \mathbf{1}=0$), then $\ker(L_s)={\rm span}(\mathbf{1})$.
Assume by contradiction that $L$ is not weight balanced, i.e., $L^\top \mathbf{y} =0$ with 
$\mathbf{y}\ne \mathbf{1}$. Then:
\begin{equation*}
    0 = \mathbf{y}^\top L^\top \mathbf{y} = \mathbf{y}^\top L_s \mathbf{y} \Longrightarrow L_s \mathbf{y} = 0 \Longrightarrow \mathbf{y} = \mathbf{1}
\end{equation*}
since ${\rm ker}(L_s)={\rm span}(\mathbf{1})$. Hence, $L$ is weight balanced.
Assume by contradiction that ${\rm corank}(L)\ne 1$, i.e., $L \mathbf{x} =0$ with $\mathbf{x}\ne \mathbf{1}$. Similarly to the previous case, it is possible to show that $L_s$ being psd of corank $1$ implies $\mathbf{x}=\mathbf{1}$. 
Since $ L_s $ psd implies $ L $ has eigenvalues with nonnegative real part, marginal stability of $ -L $ follows straightforwardly.
\qed

\medskip
\prooftx{Lemma~\ref{lemma:irreducible}}
\begin{enumerate}
\item Let $\mathcal{G}$ be a signed digraph with signed Laplacian $L$ (with at least $2$ nodes). First, we establish an equivalent condition for $\mathcal{G}$ to have a rooted spanning tree. In the unsigned graph case it has been proven, see \cite[Theorem~5]{Moreau2005Stability}, that $\mathcal{G}$ (unsigned) has a rooted spanning tree if and only if for every pair of nonempty, disjoint subsets $\mathcal{V}_1,\mathcal{V}_2\subset \mathcal{V}$ there exists a node that is an in-neighbor of $\mathcal{V}_1$ or $\mathcal{V}_2$ (see also \cite{Bullo-Lectures}). Since the proof does not depend on the specific weights on the edges of the graph, it holds also in the case of a signed graph $\mathcal{G}$.
This means that, for every pair of nonempty, disjoint subsets $\mathcal{V}_1,\mathcal{V}_2\subset \mathcal{V}$, after an adequate permutation, $L$ can be rewritten as
\begin{equation*}
  \!\!\!  L = \begin{bmatrix} L_{11} & \! \! L_{12} \\ 0 & \! \! L_{22}\end{bmatrix}\!, L_{12}\ne 0, \text{ or }\; 
    L = \begin{bmatrix} L_{11} & \! \! 0 \\ L_{21} & \! \! L_{22}\end{bmatrix}\!, L_{21}\ne 0,
\end{equation*}
assuming an edge from $\mathcal{V}_2$ to $\mathcal{V}_1$, or from $\mathcal{V}_1$ to $\mathcal{V}_2$, respectively.

Now, assume by contradiction that $L$ is of corank $1$, but $\mathcal{G}$ does not have a rooted spanning tree. Then, there exists a pair of nonempty, disjoint subsets $\mathcal{V}_1,\mathcal{V}_2\subset \mathcal{V}$, such that, after an adequate permutation, $L$ can be rewritten as
\begin{equation*}
    L = \begin{bmatrix} L_{11} & 0 \\ 0 & L_{22}\end{bmatrix}.
\end{equation*}
Then, $L \mathbf{1}=0$ implies that $0 \in \Lambda(L_{11})$, $0\in \Lambda(L_{22})$, and ${\rm ker}(L)={\rm span} \left(\begin{bmatrix}\mathbf{1}_{{\rm card}(\mathcal{V}_1)}\\0\end{bmatrix} ,\begin{bmatrix}0\\ \mathbf{1}_{{\rm card}(\mathcal{V}_2)}\end{bmatrix} \right)$. Consequently, $L$ is not of corank $1$.

\item In both statements assume, by contradiction, that $L$ is reducible, i.e., there exists a permutation matrix $P$ s.t. $P^\top L P = \begin{bmatrix} L_{11} & L_{12} \\ 0 & L_{22} \end{bmatrix}$. 
	
	\noindent
	Assume that $-L$ is EEP, i.e., $\exists\, d\ge 0$ s.t. $B=d I-L\epos 0$ (see Lemma~\ref{lemma:ev_exp_pos}). 
	Then $B$ is also reducible, since $P^\top B P = \begin{bmatrix} d I-L_{11} & -L_{12} \\ 0 & d I-L_{22} \end{bmatrix}$. It follows that $(P^\top B P)^t = \begin{bmatrix} (d I-L_{11})^t & \ast \\ 0 & (d I-L_{22})^t \end{bmatrix}$ for all $t\ge 1$, i.e., $P^\top B P$ is not EP and, consequently, $B$ is not EP.
	
	\noindent
	Assume that $L$ is weight balanced of corank $1$. Then $L\mathbf{1}=L^\top\mathbf{1}=0$ implies that $0 \in \Lambda(L_{11}^\top)=\Lambda(L_{11})$ and that $0\in \Lambda(L_{22})$. 
	Consequently, $L$ is not of corank $1$.
	\end{enumerate}
	$\hfill\square$
	
\medskip
	\prooftx{Lemma~\ref{le:nec:DT}}
	To prove the necessity, we need to prove that $\rho(W)=1$ and $1$ is a simple and strictly dominating eigenvalue. 
	As $\mathbf{x}(t)=W^t\mathbf{x}(0)$, we need to analyze $W^t$ as $t$ grows to infinity. From $ W \mathbf{1} = \mathbf{1} $, it is $ \lambda(W) =1 $, hence $\rho(W)\geq 1$. If $\rho(W)>1$, it holds $\rho(W^t)=(\rho(W))^t\to \infty$, i.e., $W^t$ diverges as $t$ grows. Therefore, $\rho(W)= 1$. 
	The rest of the proof can be obtained by an argument in the same spirit as that of Lemma~\ref{le:nec} for what concerns both the multiplicity of $ \lambda(W) =1 $ and its strict dominance. 
	The sufficiency is then obtained repeating the reasoning which led to \eqref{eq:lim}.
	$\hfill\square$

\medskip

	\prooftx{Theorem~\ref{thm:symm-sign}}
	$W$ symmetric and marginally Schur stable  with $ \lambda_1(W)=1 $ simple and strictly dominant implies $ | \lambda_1 (W)| > | \lambda_i(W)| $, $ i=2, \ldots, n $. 
	Combined with $ W \bfone=\bfone $, it means that $ W = W^\top  \in \mathcal{PF}$ or, from Theorem~\ref{thm:PFforEvPos}, that $ W$ is EDS. 
	Viceversa, $W$ EDS means $W  \stackrel{\vee}{>} 0 $, i.e., $ \rho(W) $ is a simple strictly dominant eigenvalue of $W$ with positive eigenvector. 
	Since it is also $ W \bfone = \bfone$, from Lemma~\ref{lemma:no_two_pos-eig}, it must be that $ \rho(W)=1$ is the eigenvalue of multiplicity 1 strictly dominating all other eigenvalues. Hence $ W $ is marginally Schur stable. 
	$\hfill\square$

\medskip

	\prooftx{Theorem~\ref{eq:ev-stoch-matrix}}
	$W$ ES implies that $ W \bfone = \bfone $ and, since $ W \stackrel{\vee}{>} 0 $, $ \lambda_1(W) = \rho(W) =1 $ must be a simple strictly dominant eigenvalue.
	Hence $ |\lambda_i (W) | < 1 $, $ i=2, \ldots, n$, meaning that $W$ is also marginally Schur stable. 
	As for the other direction, if $ W $ such that $ W \bfone = \bfone $ is marginally Schur stable with $ \lambda(W)=1 $ simple and strictly dominant, it means that $ | \lambda_i(W)| <1 $, $ i=2, \ldots, n$, i.e., $ \rho(W) =1 $, hence $W \in \mathcal{PF}$. 
	$\hfill\square$

\medskip
	\prooftx{Corollary~\ref{cor:weight-bal-DT}}
	(i) $ \Longleftrightarrow $ (ii) Follows combining Theorem~\ref{eq:ev-stoch-matrix} with $ W \mathbf{1} = W^\top \mathbf{1} = \mathbf{1}$. 
	
	\noindent
	(ii) $ \Longleftrightarrow $ (iii) If $W$ is normal, then it can be decomposed as $ W = U DU^\top $ where $U$ is orthonormal and $ D $ similar to \eqref{eq:D-normal-diag}, with $ \mu_1 =1 $ and $ | \mu_i |<1$, $ i=2, \ldots, \ell$, and $ \sqrt{ \nu_j^2 + \omega_j^2} <1 $, $ j=1, \ldots, \frac{n- \ell}{2}$. 
	From this, in fact, we have $ W^\top W = U (D^\top D) U^\top $ 
	, and $ I - W^\top W $ has eigenvalues $0 $, $ 1- \mu_i^2 >0$, $ i=2, \ldots, \ell$, and $ 1-( \nu_j^2 + \omega_j^2) >0 $, $ j=1, \ldots, \frac{n- \ell}{2}$, i.e., $ I - W^\top W $ is psd of corank 1. 
	$\hfill\square$
	
\section{Time-varying digraphs}
\label{app:TV}
	\prooftx{Theorem~\ref{th:form_c}}
	Denote $ \Pi = I - \mathbf{1} \mathbf{1}^\top /n $ the projection matrix onto $ {\rm span}(\mathbf{1})^\perp$. Then $ \mathbf{x}(t) $ can be split as $  \mathbf{x}(t) = \Pi  \mathbf{x}(t)  + (I - \Pi )  \mathbf{x}(t)  = \delta(t) + \alpha \mathbf{1} $, where $\alpha(t)=\frac{1}{n}\mathbf{1}^\top  \mathbf{x}(t)$ is the norm of the projection of $\mathbf{x}(t)$ onto ${\rm span}(\mathbf{1})$, and $ \delta(t) = \mathbf{x}(t) - \alpha (t) \mathbf{1} $ is the projection onto $ {\rm span}(\mathbf{1})^\perp$. 
	Here we omit the argument $t$ for simplicity of expression. 
	By \eqref{eq:form_c}, 
	\begin{equation}
		\dot{\alpha}=\frac{1}{n}\mathbf{1}^\top  \dot{\mathbf{x}}= -\frac{1}{n}\mathbf{1}^\top  L_{\sigma(t)}\mathbf{x}=0.
	\end{equation}
	The third equality is due to the weight balance of each $L_k,k\in[m]$. Therefore, $\alpha(t)$ is invariant, i.e., $\alpha(t)=\alpha(0)=\frac{1}{n}\mathbf{1}^\top  \mathbf{x}(0), t\geq 0$.
	
	As for the ``disagreement" projection $\delta$, it holds
	\begin{equation}
		\dot{\delta}=\dot{\mathbf{x}}=-L_{\sigma(t)}\mathbf{x}=-L_{\sigma(t)}\delta.
	\end{equation}
	Define a candidate Lyapunov function $V=\frac{1}{2}\|\delta\|^2\geq 0$. We have
	\begin{equation}
		\dot{V}=-\delta^\top  \frac{L_{\sigma(t)}+L_{\sigma(t)}^\top }{2}\delta.
	\end{equation}
	By Corollary \ref{cor:weight-bal1}, for all $k\in[m]$, $\frac{L_k+L_k^\top }{2}$ is psd of corank $1$, and $\mathbf{1}$ is its eigenvector corresponding to the eigenvalue $0$. Let 
	\[\lambda^\ast=\min_{k\in[m]}\lambda_2(\frac{L_k+L_k^\top }{2})>0. \] 
	$\lambda^\ast$ is real due to the symmetry of $\frac{L_k+L_k^\top }{2}$. By the Courant-Fisher theorem \cite{horn2012matrix}, it holds 
	\[\frac{\dot{V}}{\|\delta\|^2}\leq -\min_{\substack{\mathbf{v}^\top  \mathbf{1}=\mathbf{0}\\
			\|\mathbf{v}\|=1}}\mathbf{v}^\top  \frac{L_{\sigma(t)}+L_{\sigma(t)}^\top }{2}\mathbf{v}=-\lambda^\ast, \]
	i.e.,
	\[\dot{V}\leq -\lambda^\ast V.\]
	This means that $V$ decreases to $0$ exponentially at the rate $\lambda^\ast$. Therefore, for any switching signal $\sigma(t)$, $\lim_{t\to\infty} \delta(t)=0$, i.e., $\lim_{t\to\infty}\mathbf{x}(t)=\alpha(0)\mathbf{1}$. The proof is then completed.
	$\hfill\square$
	
\medskip

	\prooftx{Theorem~\ref{th:form_d}}
	The normality of $W_k$ implies that $W_k^\top  \mathbf{1}=\mathbf{1}$. Similarly to the proof of Theorem \ref{th:form_c}, let $\alpha(t)=\frac{1}{n}\mathbf{1}^\top  \mathbf{x}(t),t\in\mathbb{Z}_\geq$. It then holds
	\[\alpha(t+1)=\frac{1}{n}\mathbf{1}^\top  \mathbf{x}(t+1)=\frac{1}{n}\mathbf{1}^\top  W_{\sigma(t)}\mathbf{x}(t)=\alpha(t).\]
	Let $\delta(t)=\mathbf{x}(t)-\alpha(t) \mathbf{1}$. We then have
	\[\delta(t+1)=W_{\sigma(t)}\mathbf{x}(t)-\alpha(t+1) \mathbf{1}=W_{\sigma(t)}\delta(t).\]
	Take a candidate Lyapunov function $V(t)=\delta(t)^\top  \delta(t)$. It holds
	\begin{equation}\label{eq:def_lya2}
		V(t+1)-V(t)=-\delta(t)^\top  (I_n- W_{\sigma(t)}^\top  W_{\sigma(t)})\delta(t).
	\end{equation}
	For $k\in[m]$, since $W_k$ is normal, it can be decomposed as $W_k=UD_k U^\top $, where $D_k$ is as in \eqref{eq:D-normal-diag}:
	%
	\begin{equation*}
		D_k=
		{\small \begin{bmatrix}  \mu_1^k \\&  \!\! \ddots \\ 
				& &\!\!  \mu_\ell^k \\
				& & & \!\! \begin{array}{cc}
					\nu_{\ell+1}^k  &  \omega_{\ell+1}^k\\
					-\omega_{\ell+1}^k & \nu_{\ell+1}^k
				\end{array} \\ 
				& & & & \!\! \ddots\\ 
				& & & &  & \!\!  \begin{array}{cc}
					\nu_{\frac{n-\ell}{2}}^k  &  \omega_{\frac{n-\ell}{2}}^k \!\! \\
					-\omega_{\frac{n-\ell}{2}}^k & \nu_{\frac{n-\ell}{2}}^k \!\! 
				\end{array}
			\end{bmatrix}
		}
	\end{equation*}
	
	Here,
	\begin{equation*}
		\begin{aligned}
			\Lambda(W_k)=\{\mu^k_1=1,&\,  \mu_2^k, \dots,\mu_\ell^k, \\
			&\nu_{\ell+1}^k\pm i\omega_{\ell+1}^k,\dots,\nu_{\frac{n-\ell}{2}}^k\pm i\omega_{\frac{n-\ell}{2}}^k\}
		\end{aligned}
	\end{equation*}
	with all the $\mu^k$, $ \nu^k$ and $ \omega^k$ real, and $U$ an orthonormal matrix. 
	Therefore, $W_k^\top  W_k=U(D_k)^\top D_k U^\top $, with
	\begin{equation*}
		(D_k)^\top D_k=
		{\small \begin{bmatrix} 1 \\ & (\mu_2^k)^2 \\& &  \ddots \\ 
				& & &  (\mu_\ell^k)^2 \\
				& & & &\!\!  \eta^k_{\ell+1}I_2 \\ 
				& & & & & \ddots\\ 
				& & & & &  &  \eta^k_{\frac{n-\ell}{2}} I_2
			\end{bmatrix}
		}
	\end{equation*}
	with $ \eta^k_j = (\nu_{j}^k)^2+(\omega_{j}^k)^2$, $ j=\ell+1, \ldots, \frac{n-\ell}{2} $.
	By Corollary~\ref{cor:weight-bal-DT},
	\[|\mu_j^k|<1,\quad j=2,\dots,\ell\] 
	and 
	\[|(\nu_j^k)^2+(\omega_j^k)^2|<1, \quad j=\ell+1,\dots,\frac{n-\ell}{2}. \]
	Then, applying the Courant-Fisher theorem to \eqref{eq:def_lya2},
	\[V(t+1)-V(t)\leq -\Tilde{\lambda}V(t),\]
	where $\Tilde{\lambda}=\max_{k\in[m]}|\lambda_{n-1}(W_k)|<1$. Therefore, 
	\[V(t)\leq (1-\Tilde{\lambda})^tV(0),\]
	which yields $\lim_{t\to\infty}\delta(t)=0$. The proof is then completed.
	
	$\hfill\square$
	
\medskip

	\prooftx{Theorem~\ref{th:CLF1}}
	From the properties of $ Q$, 
	\[
	U = 
	[\begin{array}{c}
		Q  \\
		\mathbf{1}^\top/\sqrt{n} 
	\end{array}]
	\] 
	is an orthonormal change of basis, and 
	\beq
	U L_k U^\top = \left[\begin{array}{cc}
		\bar{L}_k & \mathbf{0}\\
		\mathbf{0}^\top & 0
	\end{array}\right], 
	\label{eq:ULU}
	\eeq
	where $ \bar L_k = Q L_k Q^\top  $ is the projection of $ L_k $ onto $ {\rm span}(\mathbf{1})^\perp$. Since $-L_k $ is EEP,  it follows from Corollary~\ref{cor:weight-bal1} that it is marginally stable and that its zero eigenvalue has multiplicity 1, while from \eqref{eq:ULU} it  follows that $ L_k $ and $ \bar{L}_k $ share all nonzero eigenvalues.  
	Consequently, $- \bar{L}_k$ is Hurwitz. 
	Considering the projection of $P$ on  $ {\rm span}(\mathbf{1})^\perp$, $ \bar P = Q P Q^\top $, then it is $ \bar P = \bar P^\top  \succ 0 $ and 
	\[
	- \bar L_k  \bar P - \bar P \bar L_k^\top   = 
	- Q L_k P Q^\top  - Q P L_k^\top  Q^\top 
	\]
	where we have used that by construction $ Q^\top  Q  = \Pi= I - \mathbf{1}\mathbf{1}^\top /n$, and that $ L_k \Pi = L_k$. 
	Existence of a CLF for the projections, i.e., of $ \bar P = \bar P^\top  \succ 0 $ such that $- \bar L_k  \bar P - \bar P \bar L_k^\top    \prec 0 $ $ \forall \, i $, implies that \eqref{eq:CLF1} holds, hence, analogously to the proof of Theorem~\ref{th:form_c}, that consensus is achieved for any switching signal $ \sigma(t)$.
	
	$\hfill\square$
	
\medskip
	\prooftx{Theorem~\ref{th:CLF-DT1}}
	From Corollary~\ref{cor:weight-bal-DT}, the projection of $ W_k $ onto ${\rm span} ( \mathbf{1})^\perp $, $ \bar W_k = Q W_k Q^\top $ is Schur stable. Denoting with $ \bar P = Q P Q^\top  $ the projection of $P$ on $ {\rm span} ( \mathbf{1})^\perp $, then $P\succ 0$ implies $ \bar P \succ 0 $ and 
	\[
	\bar W_k^\top \bar  P \bar W_k - \bar P =
	Q ( W_k^\top  P W_k - P)Q^\top  \prec 0, \quad k = 1, \ldots, m 
	\]
	implies \eqref{eq:CLF-DT1}.
	The equality above follows from $ \Pi = Q^\top  Q = I - \mathbf{1} \mathbf{1}^\top /n $, $ W_k\Pi  = W_k - \mathbf{1} \mathbf{1}^\top /n $ and $ Q \mathbf{1} =0 $.
	$\hfill\square$

\medskip

	\prooftx{Proposition~\ref{prop:bip}}
	Property~1 is true by construction.
	Property~2 follows from the fact that, since $ S^2 =I $, the system \eqref{eq:ode:laplacian-b} corresponds to a change of basis w.r.t. to \eqref{eq:ode:laplacian}, and consequently Property~3 follows from Lemma~\ref{le:nec}.
	$\hfill\square$

%


\end{document}